\documentclass[onecolumn]{article}

\usepackage[utf8]{inputenc}
\usepackage{lmodern}
\usepackage{authblk}
\usepackage{amsmath}
\usepackage{booktabs}
\usepackage{makecell}
\usepackage{caption}
\usepackage{graphicx,xcolor}
\usepackage[
    top=2.54cm,
    bottom=2.54cm,
    left=2.54cm,
    right=2.54cm
]{geometry}
\usepackage[normalem]{ulem}
\usepackage{placeins}               
\usepackage[
    separate-uncertainty = true,
    locale               = UK,
    group-separator      = {,},
    exponent-product     = \cdot
]{siunitx}

\usepackage[
    backend=biber,
    style=numeric,
    sorting=none,
    giveninits=true,
    maxcitenames=1,
    maxbibnames=1
]{biblatex}
\DeclareSourcemap{
  \maps[datatype=bibtex]{
    \map[overwrite]{
      \step[fieldsource=shortjournal, final]
      \step[fieldset=journaltitle, origfieldval]
    }
    \map{
      \step[fieldset=url, null]
      \step[fieldset=urldate, null]
      \step[fieldset=doi, null]
    }
  }
}

\AtEveryBibitem{%
  \clearlist{language}\clearfield{langid}%
  \clearfield{urldate}\clearfield{urlyear}\clearfield{urlmonth}\clearfield{urlday}%
  \clearfield{urlendyear}\clearfield{urlendmonth}\clearfield{urlendday}%
  \clearfield{url}\clearfield{doi}\clearfield{note}\clearfield{month}%
  \clearfield{abstract}\clearfield{keywords}\clearfield{file}%
  \clearfield{series}\clearfield{issue}%
  \clearlist{location}\clearlist{address}\clearname{editor}%
  \clearfield{issn}\clearfield{isbn}%
}

\usepackage[colorlinks=true, allcolors=blue]{hyperref}

\graphicspath{{./figures/}}

\title{\textbf{Passively stable divided-pulse spectral broadening in multi-pass cells: beam-splitter and birefringent-crystal approaches}}

\author{Nayla Jimenez$^{1,2,3,*}$,  Henrik Schygulla$^{1}$, Henrik Tünnermann$^{1}$, Arthur Schönberg$^{1}$, Meenakshi Mini Ashok$^{1,2,3}$, Nikita Khodakovskiy$^{1}$, Victor Hariton$^{1}$, Gunnar Arisholm$^{4}$,Hüseyin Çankaya$^{1}$, Ingmar Hartl$^{1}$ and Marcus Seidel$^{1,2,3,*}$}

\affil{$^1$Deutsches Elektronen-Synchrotron DESY, 22603 Hamburg, Germany}
\affil{$^2$Helmholtz-Institute Jena, 07743 Jena, Germany}
\affil{$^3$GSI Helmholtzzentrum für Schwerionenforschung GmbH, 64291 Darmstadt, Germany}
\affil{$^4$FFI (Norwegian Defence Research Establishment), Kjeller, Norway}

\affil[*]{nayla.jimenez@desy.de \\ marcus.seidel@desy.de}   

\begin{document}

    \maketitle
    
\begin{abstract}
    Nonlinear pulse compression by means of multi-pass cells has enabled ultrafast light sources with unprecedented combinations of peak and average power. We address one of the major challenges of the method, namely pulse energy up-scaling, by an experimental and numerical study of divided-pulse spectral broadening, comparing division schemes based on birefringent crystals and on 50:50 intensity beam splitters. In the first scheme, a sequence of three birefringent crystals was used to generate a train of eight pulses, each with 204\,fs duration. After spectral broadening in a bulk multi-pass cell, the pulses were recombined in another crystal sequence with 95\,\% efficiency and compressed to 43\,fs duration and 2.4\,GW peak power. In a second experiment, pulse division and recombination were implemented using 50:50 intensity beam splitters and free path delay lines. A train of eight spectrally broadened pulses was coherently combined with 82\,\% efficiency. Eleven-fold compression to 89\,fs duration and a peak power of 5.2\,GW was achieved. Despite the free path delay lines, passive stability with 0.5\,\% RMS energy fluctuations over a 12-hour measurement period was demonstrated. The sources of recombination losses and side-pulse generation were identified by numerical studies which showed  good agreement with the experiments. To investigate the potential for energy scaling, a  nonlinear compression simulation of 1\,J energy pulses from 940\,fs to 52\,fs duration by means of an only 2-m long multi-pass cell is presented. The birefringent crystal pulse division approach can be applied to nonlinear post-compression of lasers with switchable repetition rate but is practically limited to GW-level input peak powers. By contrast, the 50:50 intensity beam splitter approach offers unmatched potential for pulse energy scaling and opens up new avenues for high-average power multi-TW pulse generation.
\end{abstract}
\vspace{5 mm}

\section{Introduction}
The advancement of strong-field physics \cite{brabec_intense_2000} has been driven by the continuous evolution of ultrafast laser sources. Applications, such as high-harmonic generation \cite{lhuillier_high-order_1995}, attosecond science \cite{varillas_roadmap_2026} and relativistic nonlinear optics \cite{mourou_optics_2006}, increasingly demand a challenging combination of high peak power and high repetition rate. Ytterbium-based laser systems have emerged as the premier platform for such applications, owing to high reliability and stability, low noise  and excellent power scalability, particularly in fiber, slab and thin-disk architectures \cite{Richardson:10}. However, the pulse duration of these systems is fundamentally constrained by emission bandwidth and gain narrowing, restricting them to the few-hundred femto- to picosecond regime.

Multipass cells (MPCs) have become a central tool for nonlinear spectral broadening and post-compression of such pulses \cite{schulte_nonlinear_2016,viotti_multi-pass_2022, hanna_nonlinear_2021}, enabling efficient generation of ultrashort pulses with excellent spatial beam quality across a wide range of energies. By distributing the nonlinear interaction over many passes, MPCs allow large B-integrals while keeping peak intensities on optical components moderate. This approach has proven scalablility from  microjoule to multi-millijoule up to 200\,mJ pulse energies \cite{pfaff_nonlinear_2023}, establishing MPCs as a key technology for cutting-edge high-power ultrafast laser systems. However, the pulse energy that can be injected into a given MPC is limited in practice by peak fluence on optics and by ionization of the nonlinear medium. Increasing the cell's geometrical size relaxes these limits at the cost of complexity and footprint, which resulted in challenging cell lengths of more than 10\,m \cite{pfaff_nonlinear_2023,kaumanns_spectral_2021}. In the light of the aforementioned extreme nonlinear optics applications, peak power-scaling has been identified as one of the key directions of MPC research \cite{viotti_multi-pass_2022}. One route towards this goal is the use of advanced cell geometries. Convex--concave \cite{hariton_spectral_2023,omar_spectral_2023} and folded cells \cite{schonberg_compact_2025} are remarkably compact, yet they shift the smallest mode size from the focus to the convex or folding mirrors, which can restrict the attainable compression factor. Bow-tie and bulk-based cells \cite{heyl_high-energy_2022,kaumanns_thesis_2020,jimenez_double-stage_2025,manz_bulk-based_2025} reach high pulse energies and even few-cycle durations, but operate close to the edge of the stability range, while compressing long input pulses by large factors \cite{beaufort_towards_2026} requires hundreds to thousands of reflections. A second route employs higher-order spatial modes \cite{kaumanns_spectral_2021,koltalo_energy_2025,ierano_energy_2026}, which preserve the standard cell geometry and have been demonstrated beyond 100\,mJ, but so far provide moderate scaling factors and require for many applications lossy conversion back to the fundamental mode. A third route, divided-pulse spectral broadening \cite{stark_divided-pulse_2022},  reduces the peak power in the time domain and leaves the MPC itself unchanged. It is particularly attractive because of its high scaling potential: ideally, the pulse energy limit scales exponentially with the number of division stages, while the spectral broadening takes place in a well-studied standard Herriott-type MPC.
	
\begin{figure}[t!]
		\centering
		\includegraphics[width=0.9\linewidth]{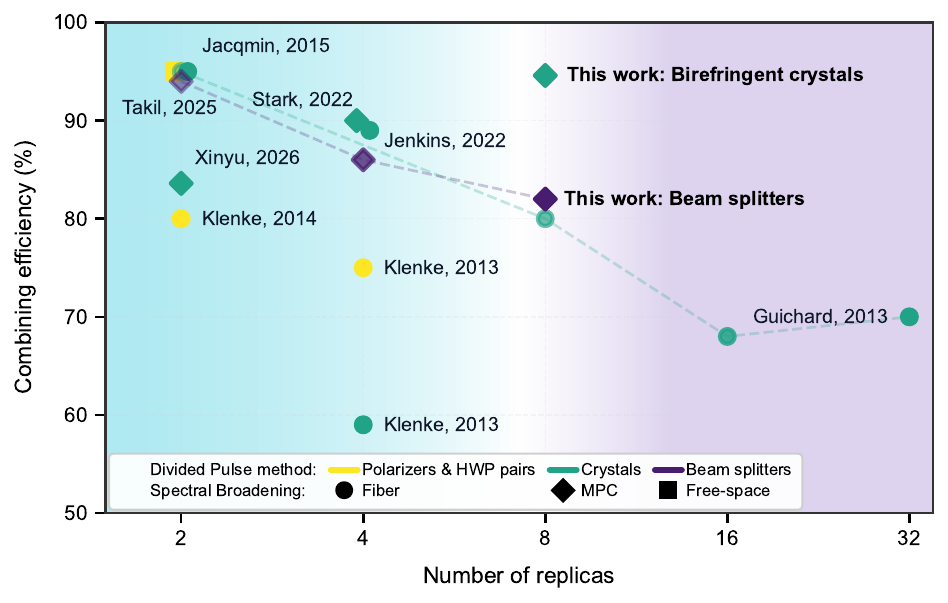}
		\caption{\textbf{Recombination efficiency vs.\ number of replicas} for divided-pulse spectral broadening schemes in the literature. Color denotes pulse division approach, marker shape the spectral-broadening method. \cite{stark_divided-pulse_2022, guichard_energy_2013, klenke_divided-pulse_2013, klenke_coherent_2014, takil_overcoming_2025,  jacqmin_passive_2015, hu_peak_2026}.}
		\label{fig:literature_summary}
	\end{figure}

Divided-pulse techniques repeatedly split a single pulse into two replicas with halved energy. The pulse train can be amplified or spectrally broadened at reduced peak intensity and subsequently recombined coherently. The approach was initially investigated in the context of pulse amplification \cite{zhou_divided-pulse_2007,kong_generation_2012} and has enabled significant energy scaling in fiber amplifiers \cite{zaouter_femtosecond_2013}. Nevertheless, chirped pulse amplification (CPA) is by far more common and divided-pulse amplification is mostly used as an extension of it. However, chirping is difficult in the context of nonlinear pulse compression because the temporal gradient of the pulse envelope determines the frequency sweep that is acquired for a given B-integral. As pulse division preserves the temporal gradient relative to the peak intensity, it is clearly better suited for spectral broadening than pre-chirping. In this context, polarizer-based pulse division schemes have been demonstrated in combination with nonlinear fibers \cite{guichard_energy_2013, klenke_divided-pulse_2013, klenke_coherent_2014} or free-space nonlinear propagation and active stabilization \cite{takil_overcoming_2025},  allowing higher pulse energies to be compressed while avoiding excessive nonlinear distortion. Alternatively, birefringent crystals were used as division and recombination units in combination with nonlinear fiber \cite{klenke_divided-pulse_2013, jacqmin_passive_2015} or an MPC \cite{stark_divided-pulse_2022,  hu_peak_2026}. Fig.~\ref{fig:literature_summary} summarizes the reported divided-pulse spectral broadening results in terms of number of divided pulses and the corresponding coherent recombination efficiencies. Division and recombination by birefrigent crystals has been the only approach to date with more than four pulses and likewise the most efficient approach enabling to recombine 95\% of the total energy for two replicas \cite{jacqmin_passive_2015}. However, the peak power-scalability of the method is severely limited. The main reason is that $>100$\,mJ pulses would require large crystal apertures in order to avoid B-integral accumulation in the crystals. Moreover, huge crystal sizes are needed because the length of the biggest crystal scales linear with the number of pulses. In addition, long crystals lead to reduced recombination efficiency because fast and slow crystal axes generally exhibit different dispersion \cite{klenke_divided-pulse_2013,guichard_analysis_2014}. Finally, the induced ps-scale delays between the pulses considerably reduce the laser-induced damage threshold intensity of the optics compared to the single pulse operation, owing to its effective pulse width dependence \cite{ristau_laser-induced_2014,astrauskas_high-energy_2017}. While these drawbacks of the birefringent crystal approach are significant, it may still be very useful for repetition rate switching of post-compressed ultrafast lasers. Therefore, we have studied the scheme and report record-high recombination efficiencies of 95\,\% for eight pulses that are spectrally broadened in an MPC. Nevertheless, the mentioned short-comings call for an alternative scheme. 
	
	Beam splitters and free-space delay lines facilitate long delays with small dispersion. This approach has been used with divided pulse amplification, but in that context,  the finite upper state lifetime and time-dependent inversion of the active material may introduce slightly varying gains and phase shifts. The consequent recombination losses can be mitigated by active path length stabilization and by fine tuning of the input power by waveplate-polarizer units. However, active control is complex, costly and causes residual noise. Moreover, it is also challenging to operate large aperture, broadband polarizers without B-integral accumulation. In divided-pulse spectral broadening, each replica of the pulse train interacts identically with the nonlinear stage, and thus the splitting and recombination units can be considerably simplified. We show that 50:50 intensity beam splitters are a viable alternative to state-of-the-art \cite{guichard_energy_2013, klenke_divided-pulse_2013, klenke_coherent_2014,takil_overcoming_2025} and easier to implement than in divided pulse amplification schemes \cite{lee_pulse_2019}. The beam splitters can be produced with large apertures, can be well dispersion-controlled over large bandwidths and require relatively simple multi-layer coatings compared to thin-film polarizers (TFPs) which relieves the coating induced stress on the substrates. They are well-suited for peak power-scaling. We achieved more than 80\,\% recombination efficiency of eight pulses, slightly exceeding the previous record efficiency for the same number of pulses \cite{guichard_energy_2013}. Furthermore, the setup exhibited excellent passive stability. The recombined pulse energies fluctuated by 0.5\,\% RMS over a 12-hour measurement period. 
	
	The paper is structured as follows. Section 2 introduces the concept and operating principle of the birefringent-crystal and beam-splitter approaches. Section 3 presents the experimental demonstrations: coherent combination of eight spectrally broadened pulses with birefringent crystals (Section 3.1) and with free-space delay lines (Section 3.2). Section 4 describes the numerical modeling of the experimental results, Section 5 provides a discussion, and Section 6 concludes the paper.

	\section{\textbf{Concept and Operating Principle}}
	
\begin{figure}[t]
		\centering
		\includegraphics[width=1\linewidth]{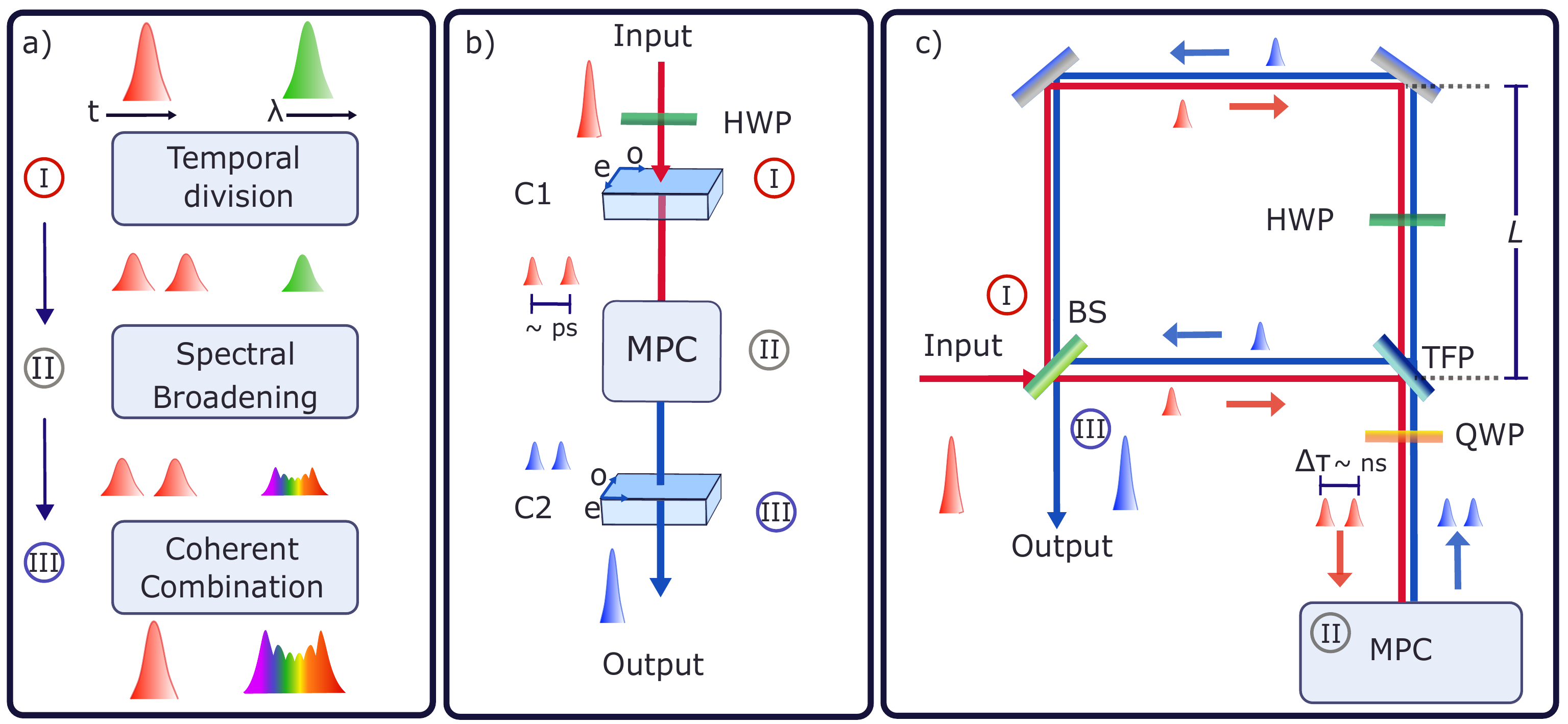}
		\caption{\textbf{a)} Principle of divided pulse spectral broadening \textbf{b)} via birefringent crystals and \textbf{c)} beam-splitter networks. C1, C2: crystals, BS: beam splitter, TFP: thin-film polarizer, HWP: half-wave plate, QWP: quarter-wave plate}
		\label{Concept}
	\end{figure}

	Both approaches rely on the temporal division of a single pulse into multiple replicas prior to nonlinear propagation in an MPC (Fig~\ref{Concept}a). The replicas originate from the same input pulse and therefore share identical spectral phase and mutual coherence at the point of division. Splitting the input pulse with peak power $P_{p,i}$ by $N_d$ dividers into $\zeta=2^{N_d}$ temporal replicas reduces the peak power ideally to $P_{p,i}/\zeta$.  Consequently, substantial pulse energy scaling directly within the existing MPC without modifying its geometry, mirror coatings, or nonlinear medium becomes feasible. Relying on the most common Herriott-type MPCs presents an important advantage of the peak power-scaling method because of their very good spatial and spatial-spectral homogeneity properties as well as the large achievable nonlinear compression factors \cite{viotti_multi-pass_2022,hanna_nonlinear_2021}. Temporal separation ensures that each replica experiences the nonlinear interaction independently, while coherent recombination after the nonlinear stage enables recovery of the single pulse energy with a (spatial) efficiency given by \cite{guichard_analysis_2014,limpert_performance_2014}:  
	\begin{equation}
		\eta = \frac{P_{\mathrm{c}}}{\sum_i P_i}
	\end{equation}
	where $P_{\mathrm{c}}$ is the combined output power (or energy), and $\sum_i P_i$ is the sum of the powers of all the individual input replicas before combining. Provided that the separation between replicas exceeds the response time of the nonlinear medium, including the delayed molecular contribution to the Kerr effect, no memory is carried over from one pulse to the next, and identical pulses acquire identical phase shift. High recombination efficiencies can therefore be achieved even after substantial nonlinear spectral broadening, provided that temporal and spatial phases of the divided pulses are matched at the recombination units and that the combiner's spectral response is flat over the broadened pulse bandwidth \cite{guichard_analysis_2014}.
	
	\subsection{\textit{Pulse division by birefringent crystals}}
	
	The birefringent crystal divided pulse scheme \cite{stark_divided-pulse_2022} (Fig~\ref{Concept}b) relies on the polarization-dependent refractive indices of anisotropic media. Either by using a half-wave plate (HWP) or by rotating the birefringent crystal optical axes by 45$^\circ$ relative to the linear polarization of the input beam, half of the pulse intensity is projected on the fast and half on the slow crystal axis. The propagation through the crystal of length $L$ results in the emergence of two orthogonally polarized replicas of the input pulse with the mutual delay 
	\begin{equation}
		\Delta \tau_\text{BC} = L\left(\left.\frac{dk_s}{d\omega}\right|_{\omega_0}-\left.\frac{dk_f}{d\omega}\right|_{\omega_0}\right ) = L\left(\frac{1}{v_{gs}}-\frac{1}{v_{gf}}\right)
	\end{equation}
	where ${dk_{s,f}}/{d\omega}|_{\omega_0}$ is the inverse of the group velocity $v_{gs,f}$
	of the pulse polarized along the slow/fast crystal axis, and $\omega_0$ is the central angular frequency of the laser pulses. The delay $\Delta \tau_{BC}$ should be large enough to avoid temporal overlap of the pulses and consequent cross-phase modulation , but also as short as possible to minimize differential dispersion between slow and fast axes \cite{klenke_divided-pulse_2013,guichard_analysis_2014} (cf. supplement~\ref{suppl:Xtal_simresults}). After spectral broadening of the divided pulses, a second birefringent crystal is used to recombine the two pulses by reversing the temporal separation and projecting them onto a common polarization axis. If the relative phase between the pulses is properly adjusted, their electric fields interfere constructively, yielding a single linearly polarized output pulse with an effectively broadened spectrum. In some earlier work \cite{guichard_energy_2013, klenke_divided-pulse_2013}, the same crystal sequence was passed a second time after spectral broadening. However, we obtained better recombination efficiencies by using a second crystal set after the MPC. By gently tilting the crystals and manipulating the extraordinary refractive indices, residual delays imprinted by dielectric coatings could be compensated. The simulations in supplement~\ref{suppl:Xtal_simresults} investigate this quantitatively.

	\subsection{Free-space delay lines by 50:50 intensity beam splitters}
	In its basic implementation (Fig~\ref{Concept}c), the input pulse is split into two replicas using a 50:50 intensity beam splitter. One replica propagates along the short path, while the second replica traverses a free-space delay line (long path) that introduces a well-defined temporal offset $\Delta \tau_\text{FS}$. The delay-line length is chosen to remain much shorter than the Rayleigh length of the collimated beam, preserving spatial mode overlap between replicas upon recombination (cf. supplement~\ref{suppl:BS_simresults}).
	
	After temporal division, the polarization of the delayed replica is rotated by  90° such that it is transmitted through the thin film polarizer (TFP), enabling spatial overlap . The replicas propagate into the multipass cell for spectral broadening. In order to recombine temporally the replicas, a reciprocal interferometric delay-line architecture is used. A quarter-wave plate (QWP) in double-pass configuration placed at the input/output of the MPC performs a 90° polarization rotation. As a result, on the return path (blue line) each replica is directed by the TFP into the complementary arm of the delay line, so it acquires exactly the delay it did not acquire on the way in. All replicas therefore accumulate the same total delay and overlap in time at the output. Temporal recombination is achieved using the same free-space delay lines employed for pulse splitting. Upon recombination at the 50:50 beam splitter, the two replicas interfere. With the same beam splitter used for division and recombination, a deviation $\delta$ from a 50:50 splitting ratio reduces the recombination efficiency only quadratically, $\eta = 1-4\delta^2$ \cite{goodno_perturbative_2010}. For a lossless beam splitter, the output fields are the  sum and difference of the partially transmitted/reflected input fields,
	
\begin{equation}
  E_{\mathrm{b}}(\textbf{r},t) = r_\text{BS}E_1(\textbf{r},t) + t_\text{BS}E_2(\textbf{r},t), \qquad
  E_{\mathrm{d}}(\textbf{r},t) = t_\text{BS}E_1(\textbf{r},t) - r_\text{BS}E_2(\textbf{r},t),
  \label{eq:bs_fields}
\end{equation}
where $E_1$ and $E_2$ are the complex fields of the two replicas after the MPC, $r_\text{BS}\approx t_\text{BS} \approx 1/\sqrt{2}$ are the complex reflection/transmission amplitudes of the beam splitter and the subscripts b and d denote the bright and dark ports, respectively \cite{saleh2019fundamentals}.  The corresponding output intensities $I_{\mathrm{b,d}}$ are
\begin{equation}
  I_{\mathrm{b,d}}(\textbf{r},t) \propto \frac{1}{2}\left(|E_1(\textbf{r},t)|^2 + |E_2(\textbf{r},t)|^2\right) \pm \mathrm{Re} \left\{E_1(\textbf{r},t)\,E_2^{*}(\textbf{r},t)\right\}.
  \label{eq:bs_intensities}
\end{equation}
 Constructive interference therefore directs the light to the bright port, while destructive interference suppresses the dark port.
	
Any perturbations slower than the return time of the pulses, e.g. mechanical vibrations, electronic noise, do not introduce  jitter or pointing between the recombined pulses. The common-path geometry of Fig.~\ref{Concept}c enables intrinsic passive stability and largely eliminates the need for active stabilization which is common for divided pulse amplification \cite{fathi_towards_2021}. However, the approach is more sensitive to slow beam drifts than the birefringent crystal method. Drifts may degrade the spatial overlap at the recombination units, and thus result in slowly decreasing recombination efficiencies. Consequently, a careful choice and engineering of optomechanics, good thermal power handling of the optics and pointing stabilization of the input beam significantly improve the long-term stability of the setup.
	
	
\section{Experimental demonstration}
	\subsection{Coherent Combination of Eight Spectrally Broadened Pulses by Birefringent Crystals}
	
	\begin{figure}[!b]
		\centering
		\includegraphics[width=0.77\linewidth]{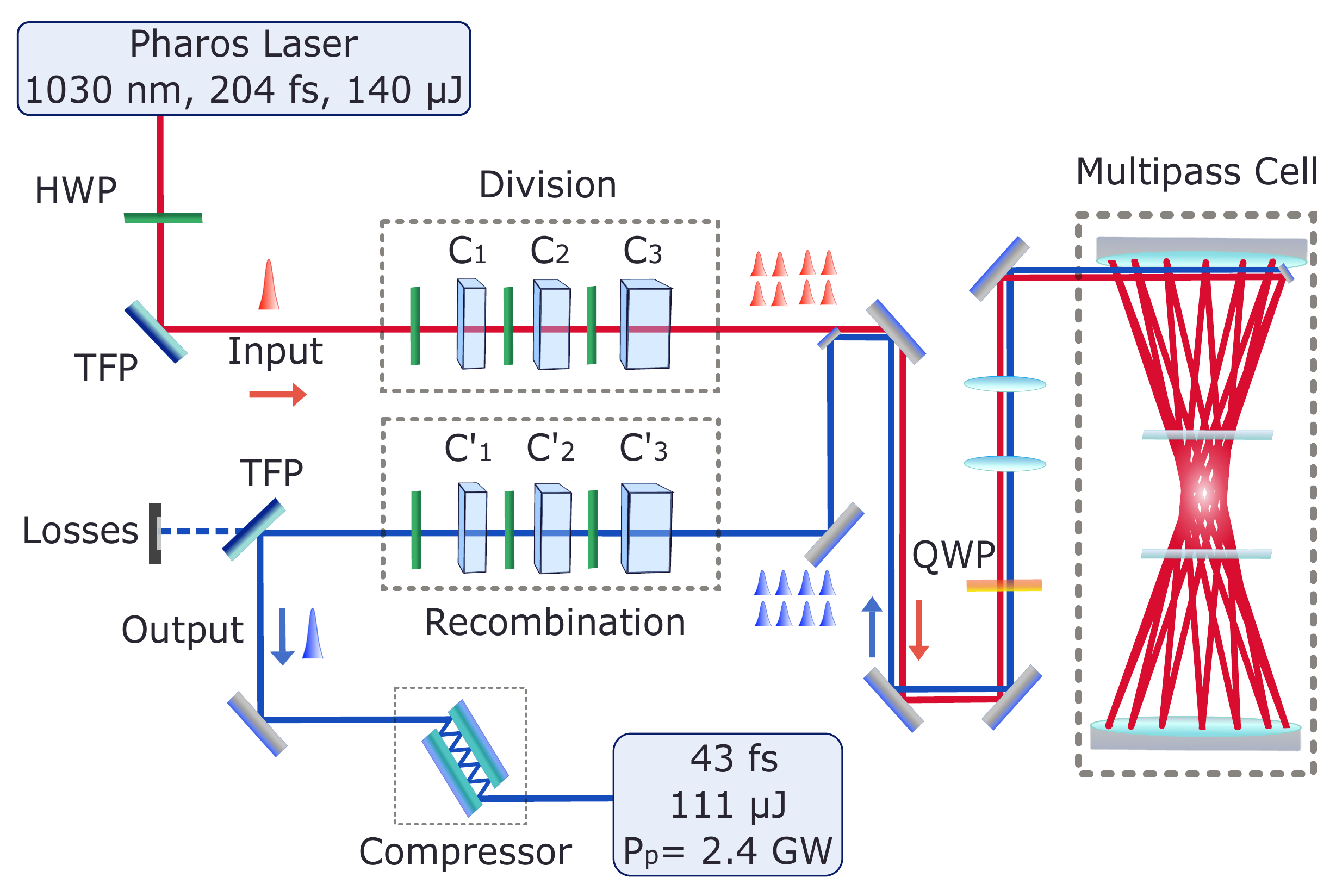}
		\caption{Schematic of the experimental setup featuring four functional stages: pulse division, MPC spectral broadening, recombination, and temporal compression.}
		\label{fig:xtal_setup}
	\end{figure}

\begin{figure}[]
		\centering
		\includegraphics[width=1\linewidth]{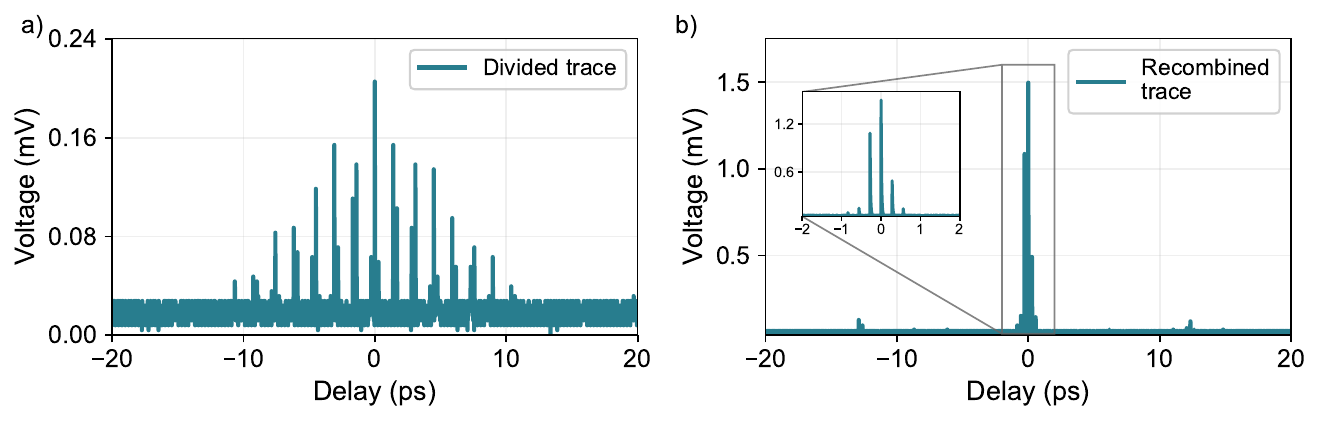}
		\caption{a) Intensity autocorrelation traces of generated pulse train. The triangular shape emerging from the convolution of a rectangular pulse train is clear. However, the set time resolution of the Femtochrome autocorrelator leads to undersampling of the individual peaks. b) Recombined output pulse. The side pulses at 12.5\,ps delay do not correspond to any crystal-induced delay and are also visible in autocorrelation measurements of the input pulses before splitting.}
		\label{ac traces.}
	\end{figure}

	The setup for divided-pulse spectral broadening based on birefringent crystals is shown in Fig.~\ref{fig:xtal_setup}. A sequence of three birefringent crystals splits the input pulse into a train of eight replicas, extending previous divided-pulse MPC experiments using four pulses \cite{stark_divided-pulse_2022, hu_peak_2026}. These results serve as a reference for the 50:50 intensity beam-splitter approach presented in the next section. As illustrated in Fig.~\ref{fig:xtal_setup}, the 204~fs input pulses with 140~$\mu$J pulse energy at a repetition rate of 7.1~kHz from a Light Conversion Pharos laser are divided into a train of eight equally spaced temporal replicas using three calcite crystals $C_1$, $C_2$, and $C_3$ arranged in order of increasing thickness ($2.5\ \text{mm}$, $5\ \text{mm}$, and $10\ \text{mm}$). The birefringence of calcite introduces a temporal walk-off of 593~fs/mm between the two axes, leading to a separation of 1.48\,ps between replicas. To achieve uniform energy distribution, a HWP placed in front of each crystal sets an equal projection of the electric field onto the crystals' fast and slow axes, ensuring balanced splitting at each individual stage. This specific arrangement also ensures that the highest-energy, undivided input pulse encounters the thinnest material ($C_1$), minimizing unwanted nonlinear interactions during the initial division stages before the per-replica energy is safely reduced. The intensity autocorrelation trace of the generated eight-pulse train is displayed in Fig.~\ref{ac traces.}a. 
	
	The spectral broadening stage consists of a 39.6-cm long MPC with two quarter-wave stack dielectric mirrors (radius of curvature \text{ROC} = -200~mm) in a 15-roundtrip Herriott-configuration ($N= 15$, $k=14$ \cite{viotti_multi-pass_2022}). Self-phase modulation is driven by two 1-mm thin, anti-reflection coated fused silica windows serving as the Kerr media. The MPC exhibits an overall transmission of 84.4\%. Since the MPC is operated close to the stability-edge, the same telescope can be used for mode-matching and collimation. A D-shaped mirror separates the incoming and outgoing beams. The beam's 45$^\circ$ angle of incidence on the dielectric mirrors (Optoman ULLM5) delay the p-polarized pulses by about 3~fs with respect to the s-polarized pulses. These delays are largely compensated by a QWP in a double-pass arrangement that rotates the field polarization by $90^\circ$. Nevertheless, the dielectric mirrors present the main source of recombination losses as explained in supplement~\ref{suppl:Xtal_simresults}.
		
	Recombination into a single pulse is performed by a second, reciprocal set of calcite crystals $C_1'$, $C_2'$, and $C_3'$, arranged in reverse order to undo the splitting sequence. An independent set of crystals was used to fine-tune the delays and to maximize the recombination efficiency.
	The autocorrelation trace of the final recombined output pulse is shown in Fig.~\ref{ac traces.}b, demonstrating a clean, single peak within the dynamic range of the autocorrelator. Following recombination, a second HWP-TFP pair performs polarization cleaning before the pulse is compressed using chirped mirrors, each providing $-200~\text{fs}^2$ of group delay dispersion (GDD), over 26 total reflections.
	The recombination efficiency of the system corresponds to 94.7\% with respect the the MPC output. The measured pulse energies at the different stages of the setup, together with the corresponding recombination efficiency, are summarized in table~\ref{tab:energy_evolution_crystals}.
	
	\begin{table}[!h]
    \centering
    \small
    \caption{Measured pulse energies at different stages of setup for eight replicas.}
    \label{tab:energy_evolution_crystals}
    \setlength{\tabcolsep}{5pt}
    \begin{tabular}{lcccc}
        \toprule
        & \textbf{MPC Input}  & 
        \textbf{After MPC} & \textbf{Recombined} & \textbf{Recombination efficiency} \\
        \midrule
        Pulse energy ($\mu$J) & 140 &  118 & 111.7 & 94.7\% \\
        \bottomrule
    \end{tabular}
\end{table}
\begin{figure}[!b]
		\centering
		\includegraphics[width=0.88\linewidth]{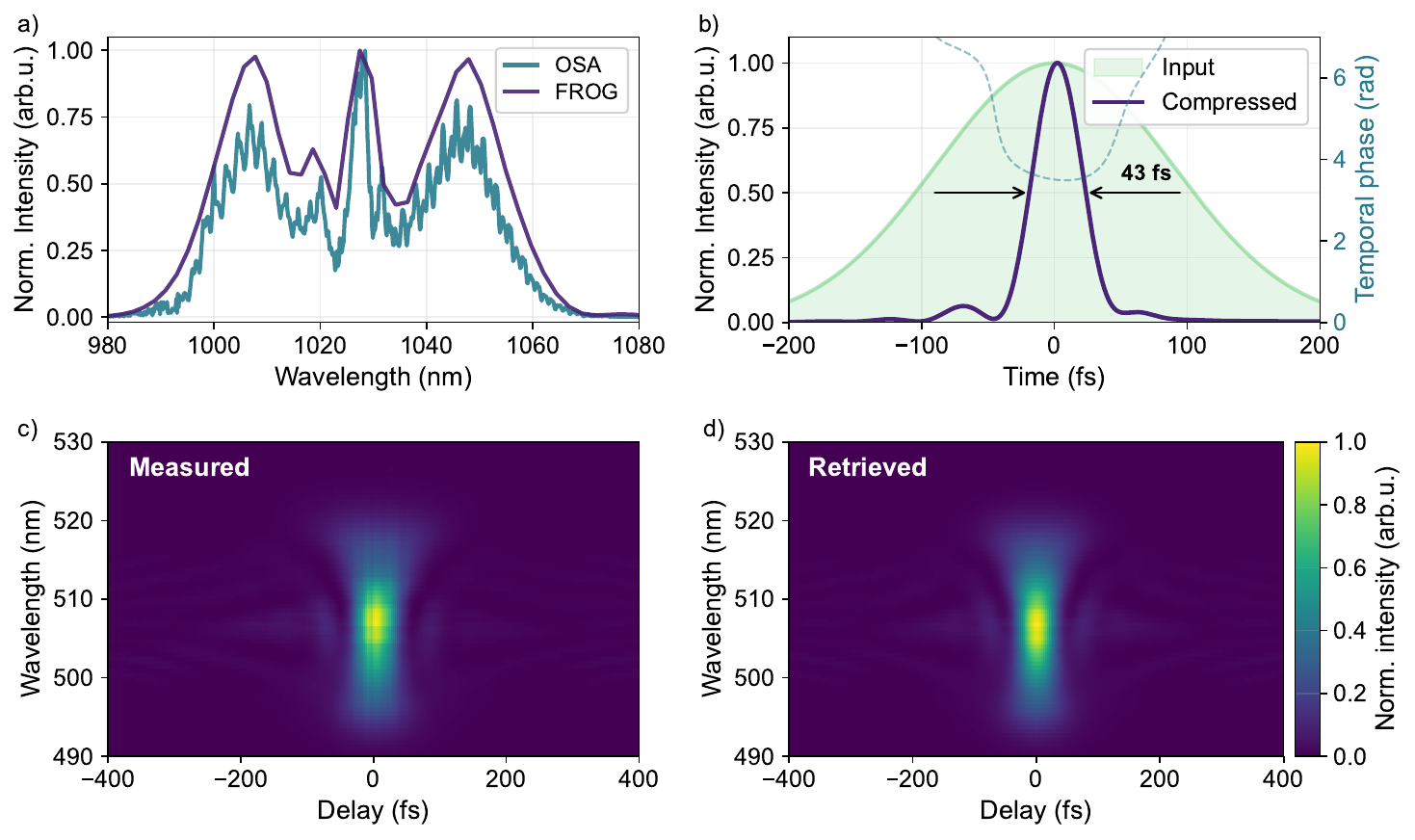}
		\caption{a) Normalized broadened spectra measured with OSA and FROG. b) Retrieved pulse by FROG
			compared to input pulse. c) Measured and d) retrieved FROG traces.}
		\label{Crystals_frog}
	\end{figure}
	
	The spectral and temporal characteristics of the compressed pulse are presented in Fig.~\ref{Crystals_frog}. The normalized broadened spectra measured via an optical spectrum analyzer (OSA) demonstrate good agreement with the data retrieved from the Frequency Resolved Optical Gating (FROG) measurement. The retrieved FROG trace and temporal profile confirm post-compression to a pulse duration of $43\ \text{fs}$. This resulted in an output pulse energy of $111.7\ \mu\text{J}$ and a peak power of $P_p = 2.4\ \text{GW}$.

\begin{figure}[]
		\centering
		\includegraphics[width=0.95\linewidth]{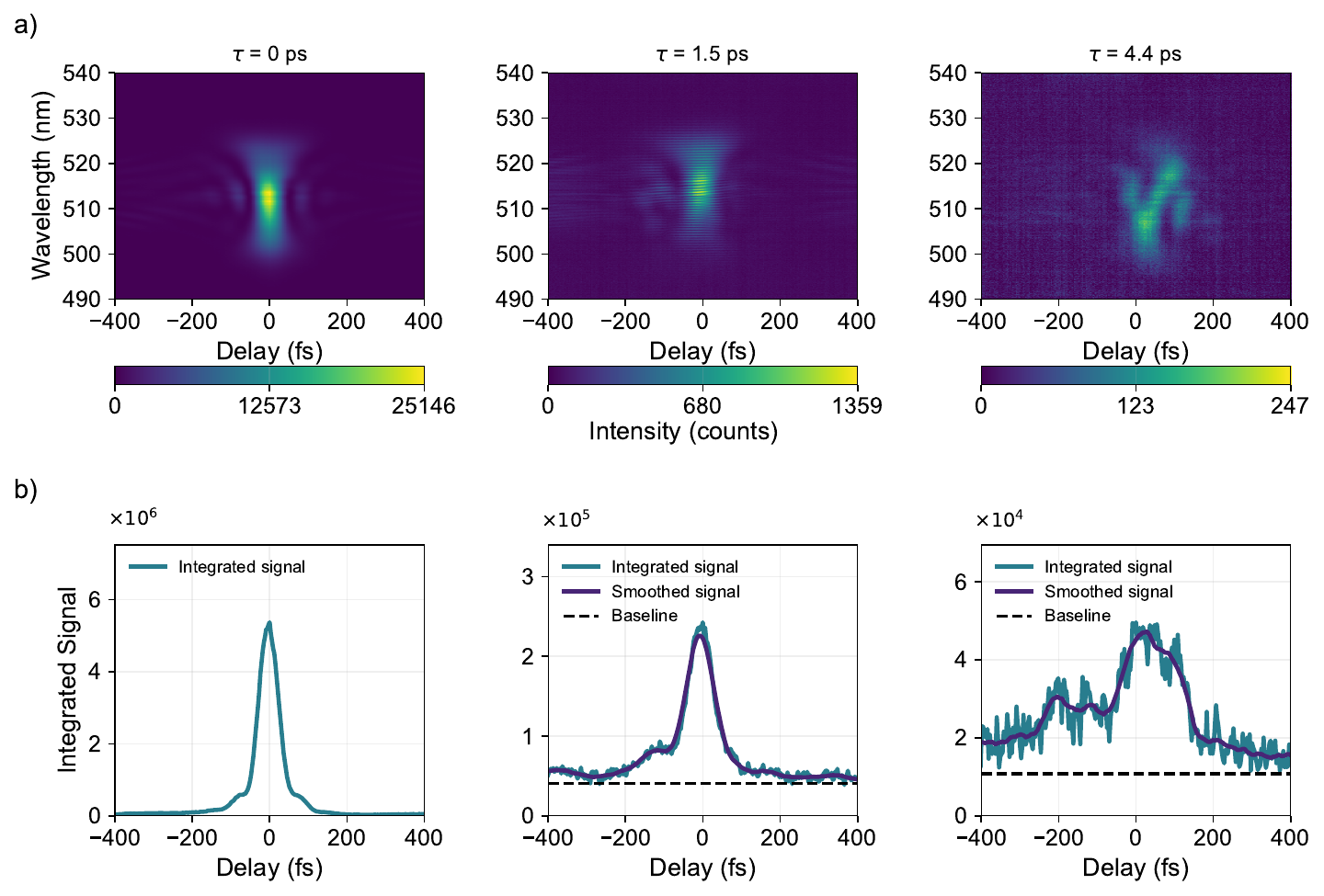}
		\caption{Temporal contrast characterization. \textbf{a)} Raw 2D FROG spectrogram capturing the main pulse and satellite pulses at 1.5~ps and 4.4~ps \textbf{b)} Corresponding wavelength-integrated temporal profile. The spectrum measurements at $\tau = 0$\,ps were done with 20\,ms integration time, the others with 200\,ms integration time.}
		\label{Side_pulses_crystals}
	\end{figure}

	To quantify the temporal contrast between the recombined pulse and its residual side pulses, direct integration of the FROG trace intensity was performed (Fig.~\ref{Side_pulses_crystals}). In this case, the wavelength-integrated FROG signal was background-corrected and integrated separately for the main peak and the side peaks at 1.5~ps and 4.4\,ps delay. After baseline and integration time correction, the strongest side-pulse at $\tau = 1.5$\,ps exhibits an amplitude of 0.82\,\% relative to the main peak. This is in good agreement with the fast modulation depth of the directly measured spectrum shown in Fig.~\ref{Crystals_frog}a. The second side pulse at $\tau = 4.4$\,ps is about an order of magnitude weaker.

\clearpage

	\subsection{Coherent Combination of Eight Spectrally Broadened Pulses by Free-Space Delay Lines }

\begin{figure}[!b]
		\centering
		\includegraphics[width=1\linewidth]{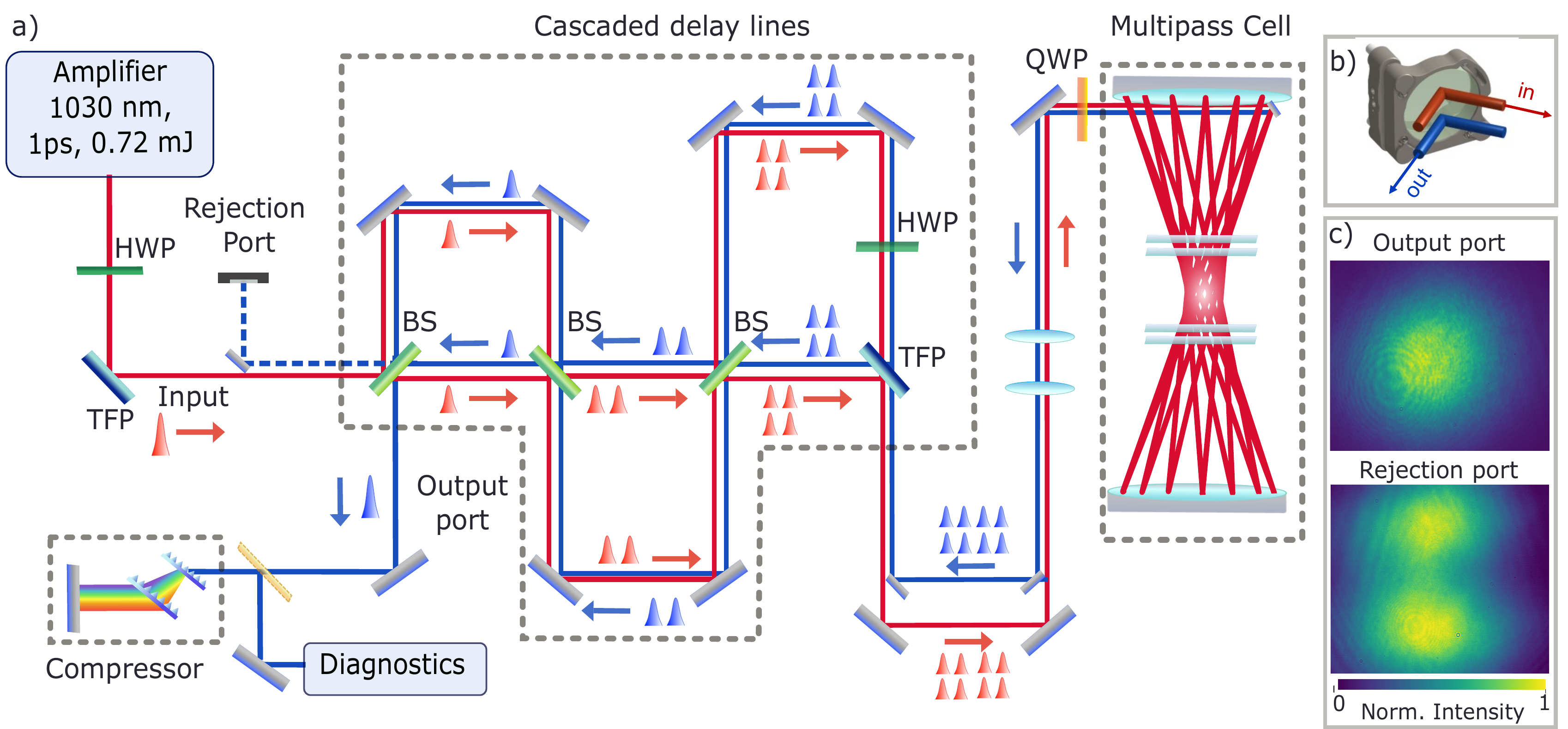}
		\caption{\textbf{Experimental layout.} \textbf{a)} Schematic of the experimental setup, comprising three cascaded delay lines for pulse division via $50:50$ beam splitters.  \textbf{b)} Mirror-reflection view of the beam paths, where the input (red) and output (blue) beams are separated by a vertical offset. c)~Beam profiles at the output and rejection ports.}
		\label{Fig 2.}
	\end{figure}

	The experimental setup shown in Fig~\ref{Fig 2.} consists of a Yb:YAG amplifier \cite{seidel_ultrafast_2022} delivering bursts at 10~Hz, with each burst comprising up to 100 pulses of 720~$\mu$J energy, 1~ps pulse duration, and an intra-burst repetition rate of 100~kHz. A HWP and a TFP are used to ensure clean s-polarization. A beam pointing stabilization unit before beam splitting is used to prevent long-term drifts. The input pulse is split into eight replicas of 90~$\mu$J each using three broadband 50:50 intensity beam splitters. The time difference $\Delta \tau_\text{FS}$ generated by the free-space delay lines corresponds to 1, 1.5 and 2~ns, creating a burst of eight pulses with equal energy content.

	\label{sec:ExpBS}

\begin{figure}[]
		\centering
		\includegraphics[width=1\linewidth]{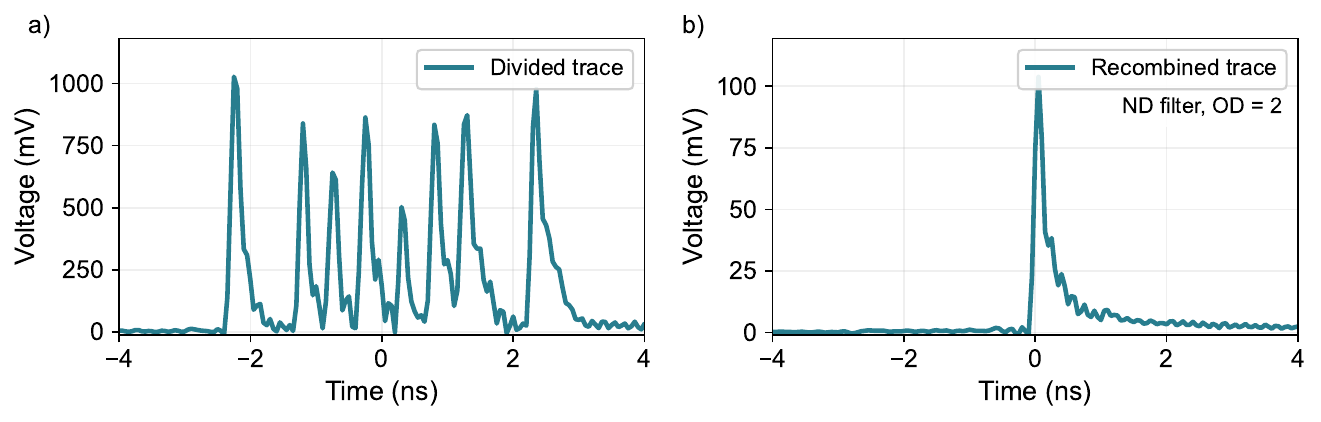}
		\caption{\textbf{Temporal characterization of the pulse division and recombination sequence.} (a) Oscilloscope trace of the generated eight-pulse train prior to injection into the multi-pass cell (MPC). (b) Recombined temporal profile measured at the output port.}
		\label{Scope_traces.}
	\end{figure}

	Fig~\ref{Scope_traces.}a shows the divided pulse train measured with a fast photodiode and a 6\,GHz bandwidth oscilloscope. The amplitude differences between replicas are attributed to the limited sampling rate of the detection system rather than the splitting itself. The delays are on the nanosecond order and can be flexibly adjusted by the distance between the beam splitters the highly reflective mirrors in the delay lines. Following the third beam splitter, a HWP rotates the reflected beam's polarization from s to p. This allows the temporally separated replicas to be spatially overlapped at a 56-degree TFP and to propagate through the nonlinear stage in a common path. A telescope then matches the beam to the MPC geometry.
	The MPC consists of two quarter-wave stack dielectric mirrors with $ROC=-200$~mm and 38.6\,cm separation. A 28-roundtrip Herriott configuration is used ($N=28$, $k=25$). Spectral broadening is driven by four 1-mm-thin anti-reflection coated fused silica plates positioned within the cavity. The MPC maintains a transmission of 85\%.
	
\begin{figure}[!b]
		\centering
		\includegraphics[width=1\linewidth]{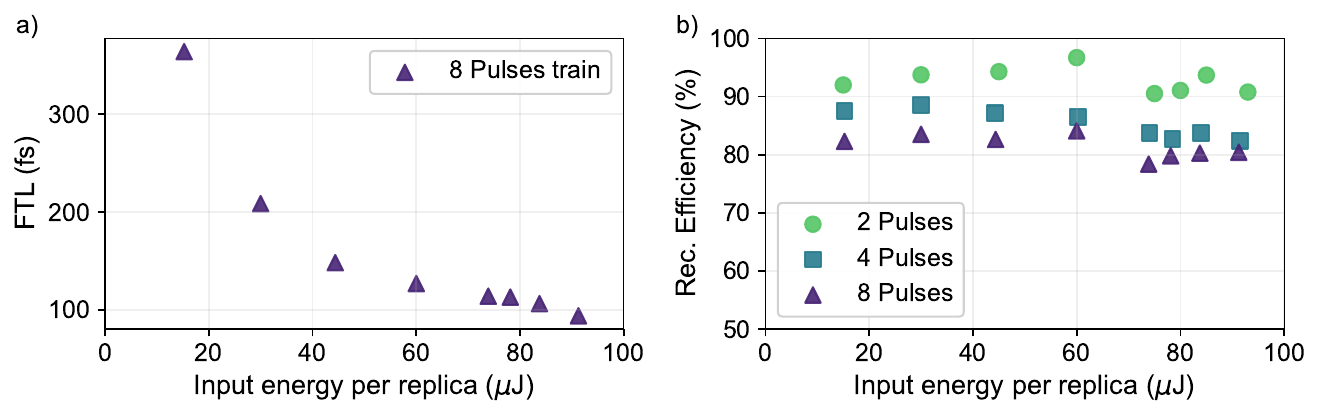}
		\caption{\textbf{Temporal compression and coherent recombination performance.} \textbf{a)} Evolution of the Fourier transform limit (FTL) as a function input energy per replica for the eight pulses case. \textbf{b)} Dependence of the recombination efficiency on the input energy per replica for two, four and eight pulses.}
		\label{Recombination.}
	\end{figure}

	To facilitate recombination, a QWP in double-pass ensures a 90$^{\circ}$ polarization rotation after the MPC. Consequently, upon exiting, each replica takes the opposite path at the TFP. By re-traversing the system in reverse, replicas that originally took the long path now take the short path on the return trip, and vice versa. This enables passive coherent recombination into a single pulse (Fig~\ref{Scope_traces.}b). The recombination efficiencies measured for two, four, and eight replicas were 93\%, 86\%, and 82\%, respectively. The measured pulse energies at the different stages of the setup for eight replicas are summarized in Table~\ref{tab:energy_evolution_bs}. The energy-scaling performance of the system is further characterized. Figure~\ref{Recombination.}(a) shows the retrieved FTL pulse duration as a function of the input energy per replica for the eight-pulse configuration. Figure~\ref{Recombination.}(b) shows the corresponding recombination efficiency as a function of the input energy per replica for two, four, and eight replicas. The recombination efficiency remains relatively stable over the investigated energy range, with higher efficiencies obtained for a lower number of replicas.
	To maximize the recombination efficiency, the spatial overlap in both the near and far fields was carefully monitored. Blocking individual delay lines allows to align the splitting and recombination units largely independently from each other. Furthermore, by introducing a vertical offset between the input and output beams, outcoupling without the need for Faraday rotators was achieved \cite{petersen_simple_2023}. 
	
		Finally, the pulses were compressed using a grating compressor in double-pass configuration with 93\% transmission efficiency. The compressed pulses were sampled and characterized by second-harmonic generation (SHG) FROG. 
		
	\begin{table}[!b]
    \centering
    \small
    \caption{Measured pulse energies at different stages of the division, spectral broadening, and recombination process for eight replicas.}
    \label{tab:energy_evolution_bs}
    \setlength{\tabcolsep}{5pt}
    \begin{tabular}{lcccc}
        \toprule
        & \textbf{MPC Input}  & 
        \textbf{After MPC} & \textbf{Recombined} & \textbf{Recombination efficiency} \\
        \midrule
        Pulse energy ($\mu$J) & 720 &  572 & 469 & 82\% \\
        \bottomrule
    \end{tabular}
\end{table}
	
\begin{figure}[]
		\centering
		\includegraphics[width=1\linewidth]{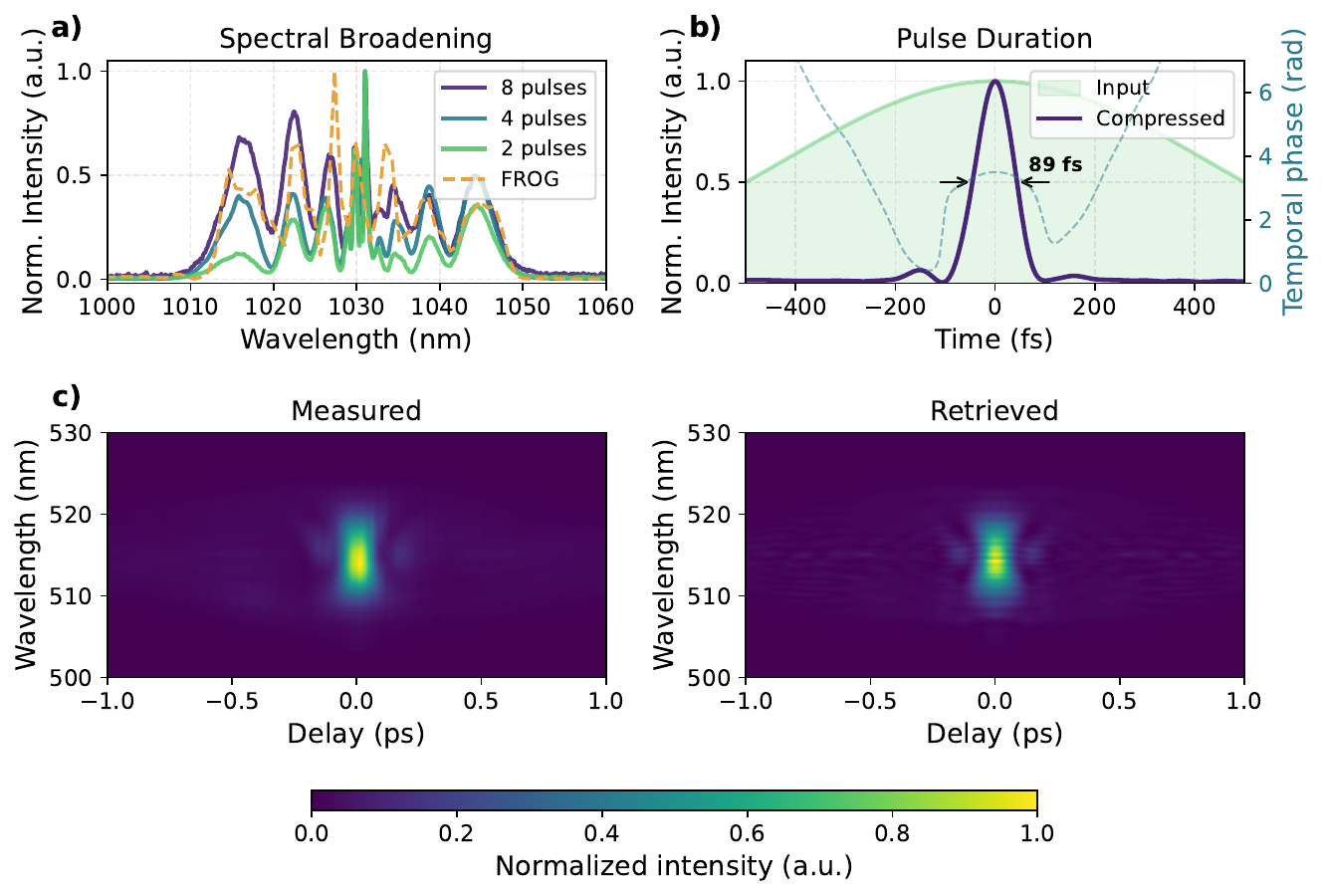}\
		\caption{\textbf{Spectral broadening and FROG characterization.} a) Normalized broadened spectra for 2, 4, and 8 pulses. (b, c) FROG characterization for the 8-pulses case: b) Retrieved recombined pulse by FROG compared to input pulse. c) Measured and retrieved FROG traces.}
		\label{BS_frog}
	\end{figure}

		The output spectra measured after recombining 2, 4, and 8 pulses exhibited nearly identical peak and valley locations, as shown in Fig~\ref{BS_frog}a, indicating that the achievable spectral broadening is essentially independent of the number of pulse replicas under the investigated conditions. This demonstrates that coherent pulse division preserves the nonlinear process performance despite the eight-fold increased input pulse energy, confirming its suitability for peak power-scaling without compromising the attainable spectral bandwidth. After optimizing the grating compressor,  SHG-FROG measurements retrieved a nearly Fourier-limited pulse duration of 89~fs with a FROG error of 0.5\%. This corresponds to a compression factor of approximately 11 and an output peak power of 5.2~GW, representing a 7.2-fold enhancement compared to the input pulse. The temporal profile of the compressed pulse is shown in Fig~\ref{BS_frog}b. 
		
\begin{figure}[!h]
		\centering
		\includegraphics[width=1\linewidth]{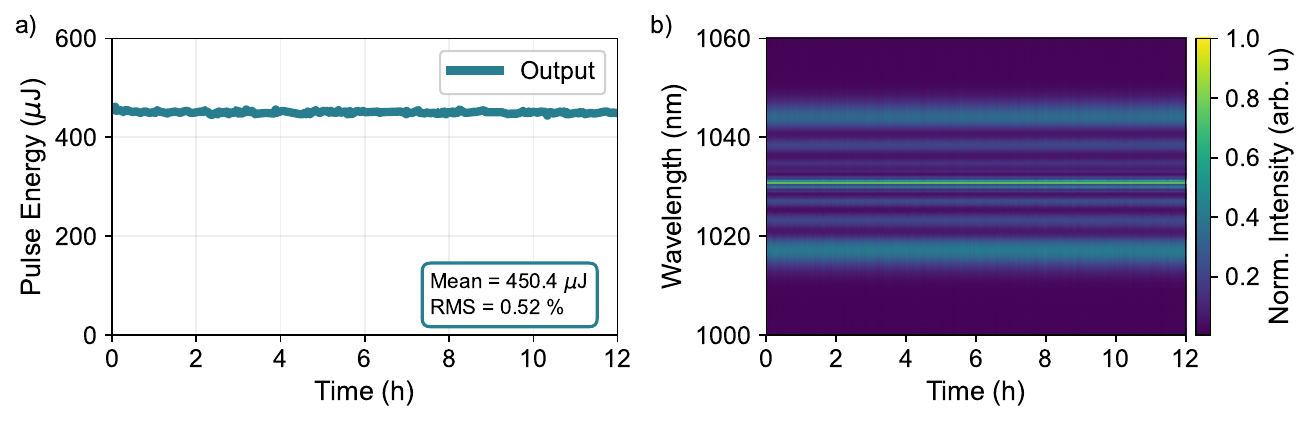}
		\caption{\textbf{Long-term stability measurements.} (a) Output pulse energy stability of the recombined pulse. (b) Spectral stability of the broadened output measured during the same interval.}
		\label{Stability}
	\end{figure}

	Long-term stability measurements were performed over 12~h of continuous operation to evaluate the robustness of the pulse replication and coherent recombination scheme Fig~\ref{Stability}. The recombined output exhibited stable operation throughout the measurement interval, with burst energy fluctuations of 0.52\% RMS. In addition, the broadened spectra remained stable over the 12-hour measurement period, indicating that also the pulse duration is stable without the need to actively control the free-space delay lines or input pulse energies.
	
	To characterize the temporal contrast of the recombined pulses, we slightly modified the SHG-FROG to exploit effective third-order nonlinear processes. The setup is described in supplement \ref{suppl:TOAC}. It could not yield an ultrahigh dynamic range map of side pulses and pedestals as known from commercial third-order autocorrelators, but provided information about the strongest pre-pulses which are most critical in ultra-high intensity applications. Breaking the time symmetry of SHG FROG and second-order autocorrelations is important for pre-pulse detection. The third-order autocorrelation signal was sufficient to saturate the spectrometer at zero-delay. Therefore, the main pulse was characterized with an 0.4 neutral density filter. After removing the filter,  pre-pulses were detected at $\Delta\tau_{FS} = -1.0\text{ ns}$ and $-1.5\text{ ns}$ with contrast ratios of $8.1 \times 10^{-4}$ and $4.2 \times 10^{-4}$, respectively. Additionally, a trailing post-pulse was resolved at $\Delta\tau_{FS} = +1.0\text{ ns}$ with a contrast ratio of $5.1 \times 10^{-4}$. The complete quantitative breakdown of the measured raw intensities and calculated contrast ratios for all pulse components is summarized in Table~\ref{tab:temporal_contrast}.

		\begin{figure}[!h]
		\centering
		\includegraphics[width=1\linewidth]{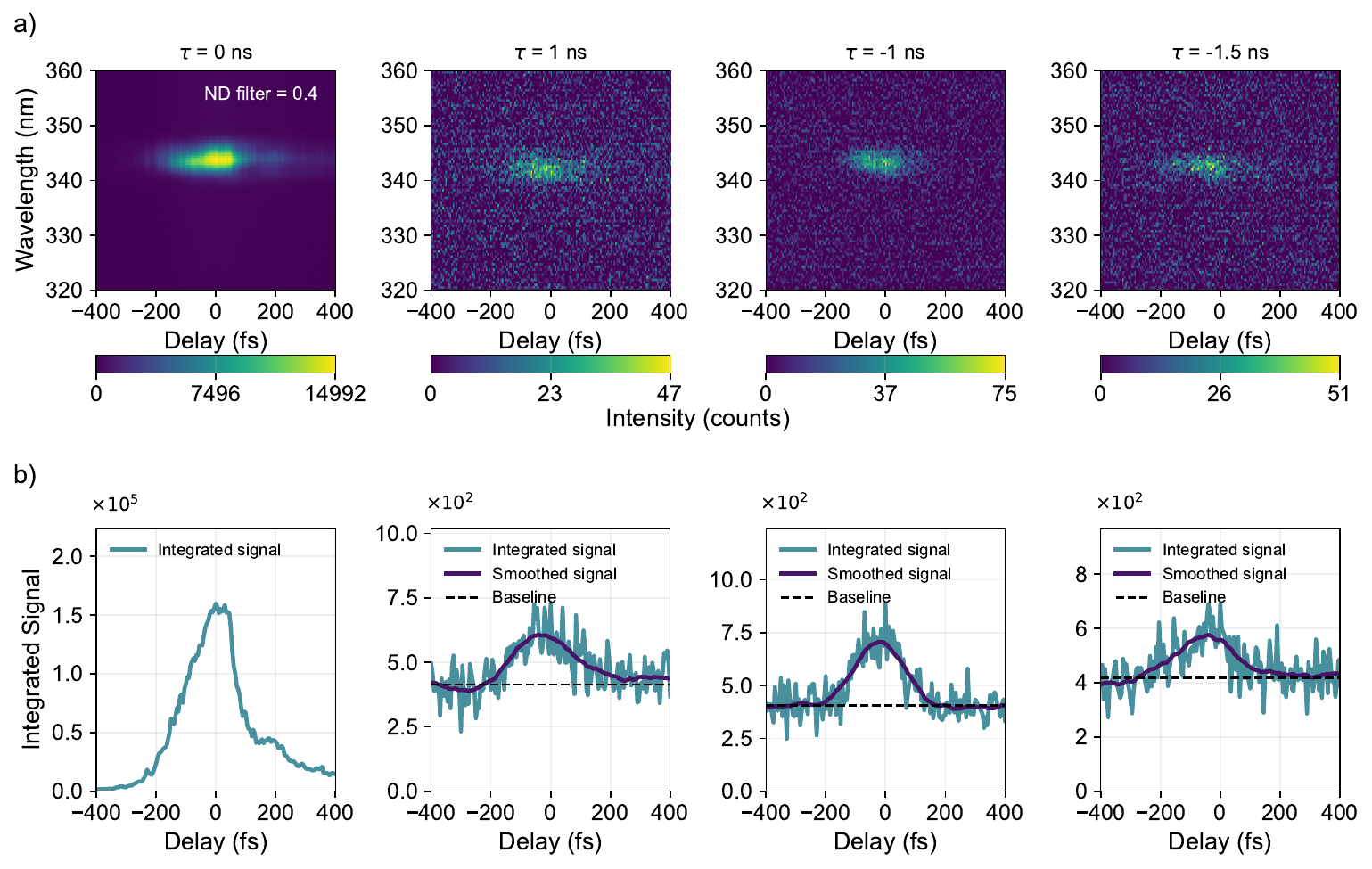}
		\caption{\textbf{Temporal contrast characterization.} \textbf{a)} Raw 2D spectrogram capturing the main pulse and temporal satellites. \textbf{b)} Corresponding wavelength-integrated temporal profile}
		\label{x-frog_bs}
	\end{figure}

	\begin{figure}[!h]
		\centering
		\small 
		\begin{minipage}[t]{0.52\columnwidth}
			\vspace{0pt} 
			\captionof{table}{Measured temporal contrast characteristics at different time delays relative to the main pulse ($t = 0$~ns). * Filter and baseline corrected data}
			\label{tab:temporal_contrast}
			\vspace{0.4em}
			\centering
			\setlength{\tabcolsep}{2.5pt} 
			\begin{tabular}{cccc}
				\toprule
				\textbf{Time Delay ($\Delta \tau$)} &  \textbf{Intensity*} & \textbf{Contrast Ratio} \\ 
				\midrule
				$+1.0$~ns & $1.92 \times 10^2$ & $5.1 \times 10^{-4}$ \\ 
				$0.0$~ns  & $3.72 \times 10^5$ & $1.00$ \\ 
				$-1.0$~ns& $3.02 \times 10^2$ & $8.1 \times 10^{-4}$ \\ 
				$-1.5$~ns & $1.59 \times 10^2$ & $4.2 \times 10^{-4}$ \\ 
				\bottomrule
			\end{tabular}
		
		\end{minipage}%
		\hfill
		\begin{minipage}[t]{0.45\columnwidth}
			\vspace{0pt} 
			\caption{Temporal contrast measured with Third-Order Autocorrelation.}
			\label{fig:xfrog_contrast}
			\vspace{0.4em}
			\centering
			\includegraphics[width=\linewidth]{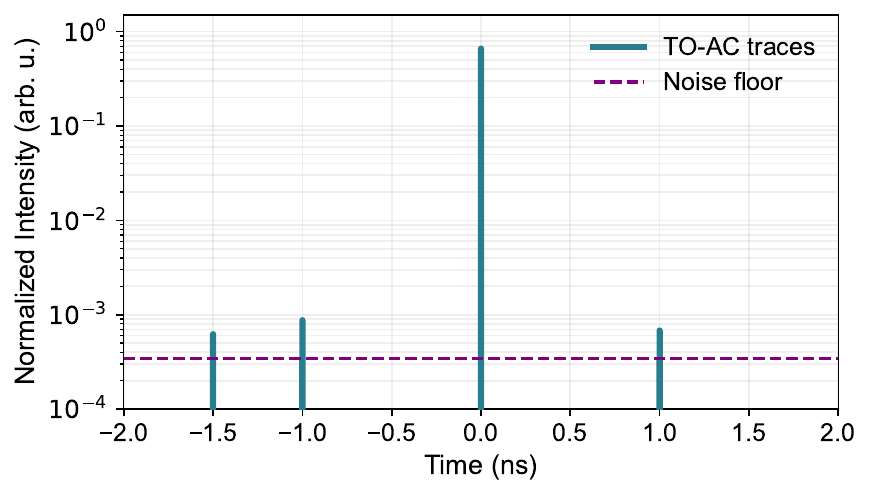}
		\end{minipage}
	\end{figure}

	\section{Numerical modeling of experimental results}
	In order to reveal the origins of recombination losses and the emergence of side pulses, we simulated the experimentally studied divided pulse spectral broadening schemes. The numerical methods are explained in the supplements~\ref{suppl:Xtal_methods} and \ref{suppl:BS_methods}. 
	\subsection{Simulations of divided pulse spectral broadening with birefringent crystals}
\begin{figure}[!b]
		\centering
		\includegraphics[width=1\linewidth]{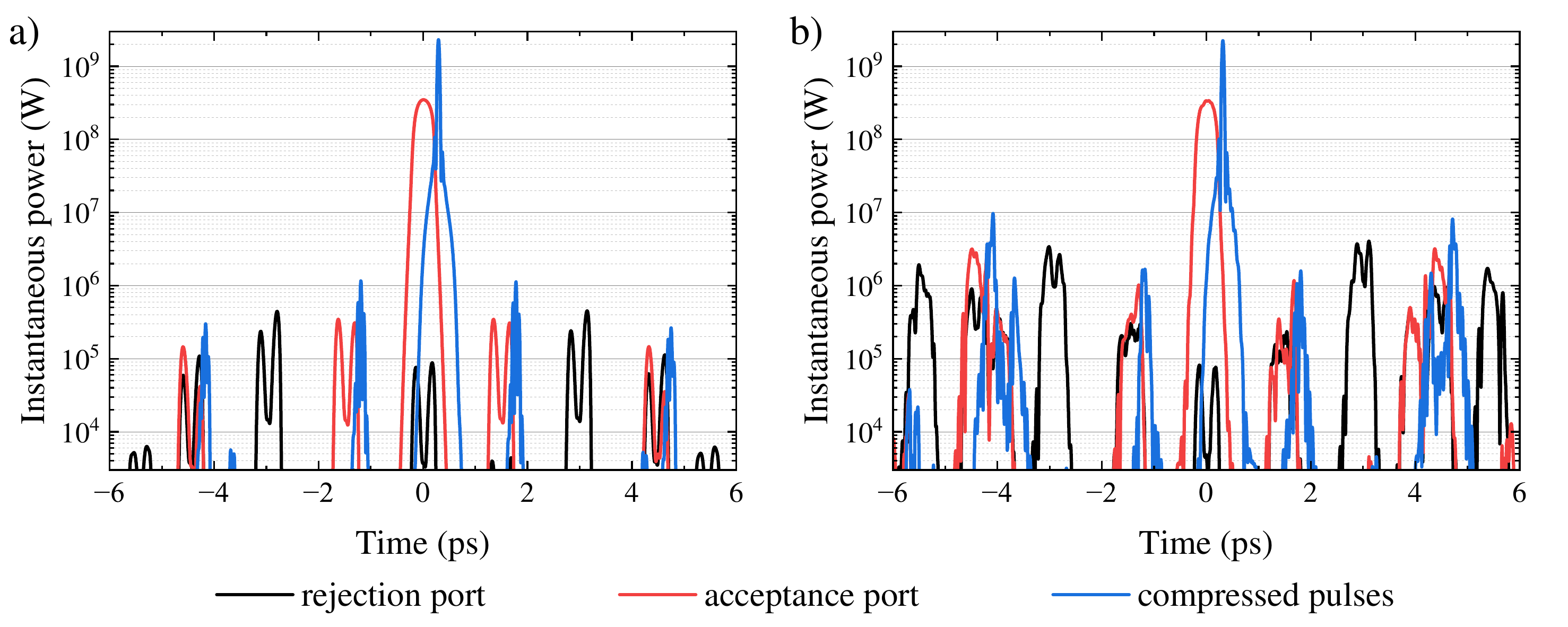}
		\caption{\textbf{Divided pulse spectral broadening simulations with birefringent crystals.} Simulated pulses for (a) ideal optics and Gaussian pulses and (b) after including the phase shifts induced by highly reflective mirrors.}
		\label{fig:Xtal_sim}
	\end{figure}

	Fig.~\ref{fig:Xtal_sim}a shows the expected output pulses for idealized optics: First, the crystal pairs have exactly the specified lengths of 2.5\,mm, 5\,mm and 10\,mm. Second, the quartz crystal thickness leads exactly to the zeroth order $\pi/2$ or $\pi$ retardance. Dispersion of the quartz waveplates and the calcite crystals is considered and computed from the Sellmeier equations derived by Ghosh et al.
		
	\cite{ghosh_dispersion-equation_1999,polyanskiy_refractiveindexinfo_2024}. Third, the polarizer behind the recombination crystals has infinite contrast and no dispersion. Fourth, no other optics are in the setup which change the relative phases between p- and s-polarized light. Under these conditions, the recombination losses are 0.17\,\% and the contrast after compression with chirped mirrors between the strongest side pulse and the main pulse is $5\times 10^{-4}$. The simulations qualitatively reproduce experimental observations (Fig.\ref{fig:Xtal_sim}): The strongest side pulse is located at the single mutual split pulse delay $\Delta\tau_{BC}$, the side-pulses at $2\Delta\tau_{BC}$ are not present in the acceptance port and the side-pulses at $3 \Delta\tau_{BC}$ are about an order of magnitude weaker than the side pulses at $\Delta\tau_{BC}$. The side pulses and limited recombination efficiencies emerge from different group velocity dispersion (GVD) along the slow and the fast axes of calcite as pointed out in earlier works \cite{guichard_analysis_2014,jacqmin_temporal_2015}. The shape of the uncompressed pulses reveals this effect. The double-peak structure indicates that the spectral wings, which were generated in the MPC, are not perfectly matched in phase upon recombination. The central frequencies, which were already present in the pulses during splitting, recombine more efficiently. Jacqmin and coworkers suggested to use additional crystals (LiNbO$_3$) in order to compensate for the differential GVD to further improve recombination efficiency and contrast \cite{jacqmin_temporal_2015}. However, our experimental results are about an order of magnitude worse than the idealized optics simulation shown in Fig.~\ref{fig:Xtal_sim}a. According to our simulations with more realistic optics, the highly reflective (HR) mirrors in our setup cause this degradation of recombination efficiency and contrast. For an oblique angle of incidence, the p-polarized pulses effectively penetrate deeper into the dielectric mirrors than the s-polarized pulses because of their lower Fresnel reflectance at each layer interface. We used HR mirrors (Optoman ULLM5) which induce a group delay difference of 3.0\,fs at 1030\,nm and 45$^\circ$ angle of incidence. To cancel this effect, polarization rotation by the QWP shown in Fig.~\ref{fig:xtal_setup} was essential. Nevertheless, there were still two HR mirrors between QWP and MPC in the setup causing linear ellipse rotation and subsequently nonlinear ellipse rotation in the MPC. By adjusting the QWP angle, this effect could be largely but not fully compensated (cf. supplement~\ref{suppl:Xtal_simresults}). Secondly, there were also two HR mirrors in the path between the QWP and the recombination crystals which were not in the path between the splitting crystals and the QWP. The resulting differential delay of about 6\,fs was compensated by slightly adjusting the angle of incidence on a recombination crystal to change its extraordinary refractive index. While the major impact from the 1.8 optical cycle delay on the recombination efficiency was canceled in this way, some excess losses and reduced pulse contrast remained as Fig.~\ref{fig:Xtal_sim}b shows. After accounting for both effects, we numerically obtained 4.2\,\% losses and a contrast of $4\times 10^{-3}$ between the main and the strongest side pulse. Both values match our experimental results.  Also the observation that the side pulse at $\Delta\tau=1.5$\,ps is nearly compressed whereas the side pulse at $\Delta\tau=4.4$\,ps remains stretched agrees with the FROG results presented in Fig.~\ref{Side_pulses_crystals}. The only qualitative difference between the numerical and the experimental results is the location of the strongest side peak which is at 1.5\,ps in the experiment and at 4.4\,ps in the simulations after introducing the differential group delay of the HR mirrors. An extended investigation of sources for recombination losses and side-pulse generation is presented in supplement~\ref{suppl:Xtal_simresults}.

	\subsection{Simulations of divided pulse spectral broadening with intensity beam splitters}
	\label{sec:simBS}
\begin{figure}[!b]
		\centering
		\includegraphics[width=1\linewidth]{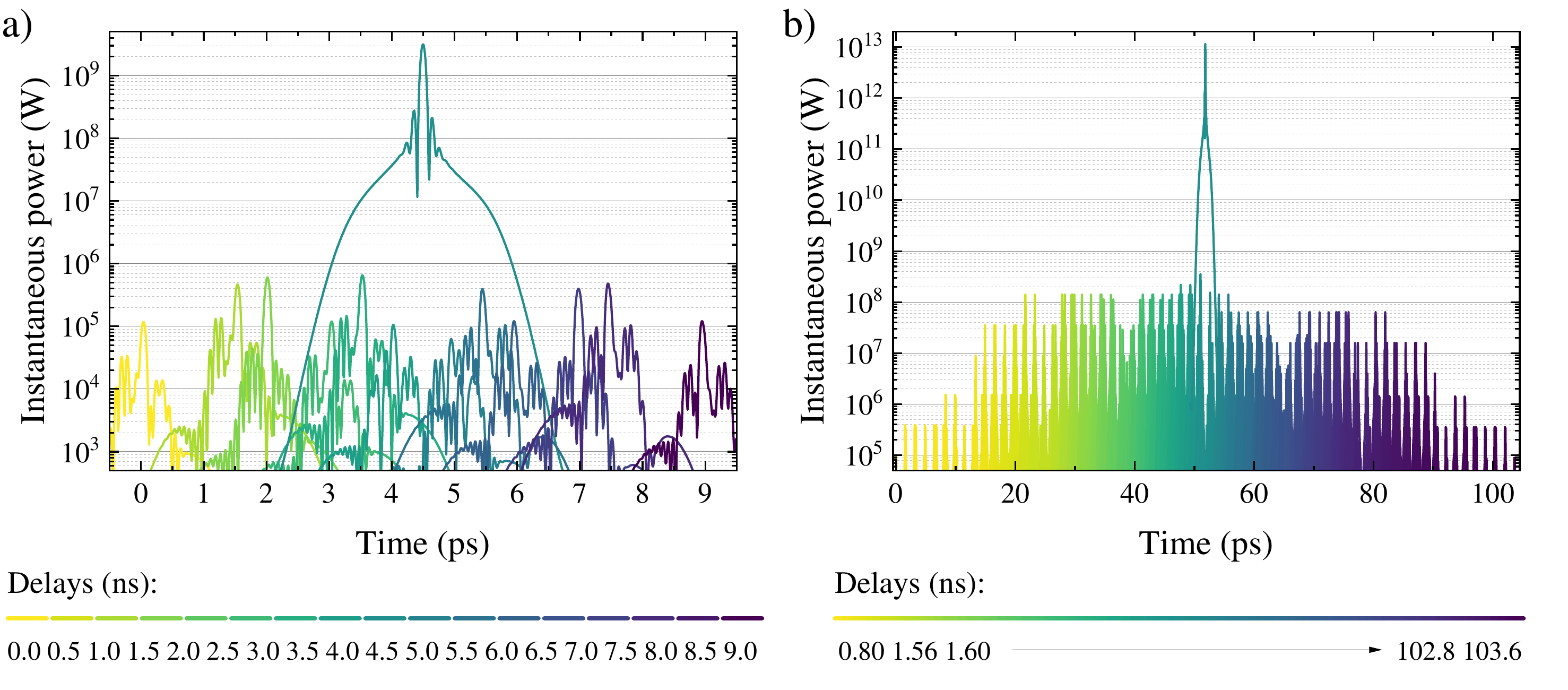}
		\caption{\textbf{Divided pulse spectral broadening simulations with intensity beam splitters.} Simulated pulses for (a) the experimental settings (b) 1\,J, 1\,TW input pulses and seven beam splitters. To visualize the side pulses and details of the main pulse, all simulated delays are divided by 1000 for plotting, i.e. the pulse do not overlap in time.}
		\label{fig:BS_sim}
	\end{figure}

	Fig.~\ref{fig:BS_sim}a shows the simulation of the divided pulse spectral broadening experiment with the 50:50 intensity beam splitters. The simulation includes linear propagation through the splitter / divider units and the nonlinear propagation through the MPC (cf. supplement~\ref{suppl:BS_methods}). Similar to the experimental findings, a side-to-main-pulse contrast on the order $10^{-4}$ is computed. However, the simulated recombination efficiency of $99.9$\,\% is much higher than the experimentally achieved one. The beam profile of the rejection port in Fig.~\ref{Fig 2.} reveals the potential reason for it. The simulation assumes cylindrical symmetry of the optical setup and perfectly flat surfaces of all reflectors. This contradicts the experimental observations. The rejection port profile of Fig~\ref{Fig 2.} shows that mainly the beam wings in the vertical direction could not be recombined with high efficiency. We attribute this to wavefront distortions the mirrors and beam splitters imprinted on the laser beam. The setscrews apply pressure to the top of the substrates. Therefore, distortions in the vertical directions are plausible. The beam splitter substrates were only 4\,mm thick, and thus vulnerable to small surface deformations. The determined contrasts of simulation and experiment are still on the same order because the non-recombined parts diverge differently from the recombined ones, and thus may not lead to a FROG signal. A detailed investigation of some contrast degradation sources is presented in supplement \ref{suppl:BS_simresults}. The simulations revealed that both the dispersion and the wavefront differences between long and short paths yield only minor contributions to the pulse contrast. The optics used in the experiments were considered in the simulations as much as the available data sheets permitted. This included beam splitter reflectance and GDD, TFP reflectance, HR mirror differential group delay and silica dispersion. The waveplates were modeled as true zero-order while compound zero-order ones were used.  Unavailable group delays, reflection and transmission curves were omitted in the simulations. We can qualitatively estimate their impact on the pulse contrast though. In particular, the back surface reflectance of the beam splitters was not precisely known. We expect that the reflections also cause side pulses with relative peak intensity on the $10^{-4}$-order as antireflection coatings typically reach transmissions of $>$99.9\,\% for silica substrates.\par

	To numerically investigate the scalability of our approach, we simulated spectral broadening of Gaussian pulses with 1\,J energy, 940\,fs duration and 1\,TW peak power. The MPC geometry from ref.~\cite{rajhans_post-compression_2023} was adapted which has recently been applied to relativistic electron acceleration with 10\,mJ-level pulses \cite{farace_multi-kilohertz_2026}. To boost the input peak power by two orders of magnitude, pulse division and recombination with seven beam splitters was simulated. We assumed the use of 6" (152.4-mm) diameter optics and an input beam diameter of 54\,mm. The large beam sizes are required to keep the B-integral in the beam splitter substrates low. As the intensity halves after each divider, demagnifying telescopes after multiple divisions can reduce the setup size as well as cost and give more flexibility in setting the delays but may cause additional losses. Additionally, pre-chirped pulses, which are only compressed while passing the MPC, present an interesting approach to avoid B-integral issues and the need for very large-aperture optics. In the simulation presented here, transform-limited Gaussian pulses were used and it is assumed that the beam size is only reduced by the mode-matching lens for the MPC. The least delay line path length is hence determined by the size of the 6"-diameter beam splitters. The choice of the maximum distance between beam splitter and HR mirror in the delay lines (that is L in Fig.~\ref{Concept}c) depends mainly on the available space in the laboratory. We have chosen to match it with the length of the spectral broadening stage, i.e. have set it to 2\,m which corresponds to the MPC mirror spacing taken from ref.~\cite{rajhans_post-compression_2023}. The remaining five delay lines were set to minimize the risk of MPC mirror damage as explained in supplement~\ref{suppl:delay_lines}. The last pulse of the train is delayed by about 52\,ns from the first. The mutual pulse delays are not equal but at least 200\,ps (cf. supplemental Fig.~\ref{fig:pulse_train}). Fig.~\ref{fig:BS_sim}b shows the simulated ensemble of pulses after recombiniation and compression. There is a broad background of more than 2000 side pulses. However, the main energy is almost fully concentrated in the main pulse and the contrast between strongest side pulse and main pulse is comparable to the simulations with three beam splitters only. This is plausible because the non-recombined side pulses get split further at beam splitters which recombine the main pulse. The central delay of about 50\,ns, i.e. the delay difference between the shortest and the longest path through the delay lines, is more than an order of magnitude longer than in the experimentally studied setup with eight pulses. As the Rayleigh length of the expanded beam is likewise increased, this does not significantly increase the side pulse amplitude. Moreover, even 50\,ns delays are too fast to introduce noise which could corrupt the passive setup stability. The simulation predicts compression to 52\,fs FWHM duration in good agreement with the 10\,mJ-level single pulses and the same MPC configuration \cite{rajhans_post-compression_2023}. In the simulations presented here, the input peak power is higher than in the previously reported experiment which is balanced by the use of circular polarization. Primarily, the numerical study indicates that well-established and highly engineered Herriott-type MPCs can be used with orders-of-magnitude higher pulse peak powers and without the need of extreme length scaling. 

	\section{Discussion}

	We have experimentally demonstrated two divided pulse spectral broadening schemes with trains of eight pulses. One is based on birefringent crystals; the other one is based on free space optical delay lines and intensity beam splitters. Both methods are complementary to each other. The birefringent crystal approach is less alignment sensitive and more robust against slow drifts. Moreover, excellent recombination efficiencies of 95\% were obtained. In combination with spectral broadening in hollow core capillaries, pulse energies of up to 10\,mJ were demonstrated with splitting and recombination by birefringent crystals \cite{jenkins_energy_2022}.
	
	Nevertheless, the peak power scalability is clearly inferior to the intensity beam splitter approach for the four reasons mentioned in the introduction section. Since free-space delays are flexible in length and nearly dispersion-less, it is straightforward to add additional splitting units and in contrast to the birefringent crystal approach \cite{guichard_analysis_2014}, there is no trade-off between number of pulses and differential dispersion.  The reported experiments showed a recombination efficiency difference between the four- and eight-pulse setup of only 4\%. Therefore, it is expected that using 16 or more pulses is feasible with only moderate additional losses. 
	
In this regard, the peak power-scaling potential of the divided pulse approach is exceptional compared to the other MPC energy scaling methods reported so-far. How large the scaling factor for a given MPC geometry and number of divided pulses is, depends on several parameters. The pulse energy in an MPC is generally limited by ionization and mirror damage. The multi-photon ionization probability of gases depends only on the pulse intensity, not on the pulse duration. Therefore, we expect that the ionization threshold scales inversely proportional to the number of pulses, i.e. increases exponentially with the number of beam splitters. Ionization was in-fact limiting the operation of the MPC with the highest reported pulse energy today \cite{pfaff_nonlinear_2023}. If the MPC input gets limited by the damage threshold of the mirrors, the energy scaling with the number of pulses may become less favorable. The last pulse of the train divided by the intensity beam splitters was delayed by 4.5\,ns from the first. Under the assumption of the common $F_\text{LIDT} \propto \tau_\text{max}^{1/2}$ relation \cite{ristau_laser-induced_2014}, where $F_\text{LIDT}$ is the laser-induced damage threshold fluence and $\tau_\text{max}$ the maximal delay, an $F_\text{LIDT}$-scaling factor of $(4.5\,\text{ns} / 1\,\text{ps})^{1/2}\approx 67$ is expected, i.e. up to six pulse dividers can be used. The simulation presented in section~\ref{sec:simBS} demonstrated that even 50\,ns maximal delays are feasible without significant recombination losses and contrast degradation. This enabled division into 128 replicas and spectral broadening of Joule-class pulses in an only 2-m long MPC which corresponds to two orders of magnitude energy-scaling. For comparison, with the birefringent crystal method, the maximal delay was only 10.5\,ps. For such durations, $F_\text{LIDT} \propto \tau_\text{max}^{0.3 – 0.5}$ is expected \cite{ristau_laser-induced_2014}, resulting in a scaling-factor between 3.2 and 6.8. Therefore, dividing into more than eight pulses with birefringent crystals would only make sense if gas ionization is the main limitation of the spectral broadening unit.
	
	In addition to energy-scaling, further pulse shortening can enhance the output peak power of the post-compressed pulses. The larger the bandwidth difference between the input and the spectrally broadened pulses, the more detrimental the differential dispersion effect becomes for the birefringent crystal approach \cite{guichard_analysis_2014}. Our simulations predict, however, that even down to 20-fs duration, the recombination losses can be kept below 10\% and even few-cycle pulse generation has already been demonstrated \cite{jacqmin_passive_2015}. The setup with the free path delay lines is currently limited by the TFP bandwidth. Using a TFP designed for 70$^\circ$ angle of incidence, as done for the 1\,J-pulse simulations, will reduce the pulse duration limit down to about 30\,fs. Alternatively, both pulse trains emerging from the last 50:50 beam splitters can counter-propagate through the MPC. Without the polarization-based combiner, even compression down to the few-cycle regime gets possible. The major drawback of such a configuration is that a height-offset between incoming and outgoing beam cannot be implemented. Therefore, the rejected light could propagate back to the laser amplifier which can be significant pulse energy in case of misalignment. Consequently, strong optical isolation becomes necessary to mitigate the risk of laser damage.
	
	We consider different applications for the birefringent and the free path delay line approaches. The birefringent crystal approach can be applied to repetition rate switching of Yb-ion lasers. Many commercial systems offer to exchange repetition rate and pulse energy while keeping the output power constant. For each setting, one would need to adjust the MPC in order to preserve pulse compression. Inserting birefringent crystals before and after the spectral broadening stage is a simple method to adjust a single MPC to the laser amplifier. For example, pump-probe experiments at the free-electron laser facility FLASH run either at 100-kHz or 1-MHz intra-burst repetition rate. We previously studied if the MPC adjustment can solely be done by gas pressure tuning \cite{viotti_flash_2023}. However, the tenfold pressure difference resulted in beam misalignment. Therefore, only a slight pressure tuning and the use of divided pulse spectral broadening can be a viable alternative. We have tested to mount the birefringent crystals on magnetic flip mounts and we observed that the divided pulse spectral broadening setup remained aligned after removing and reinserting the birefringent crystal pairs. The free path delay line approach can be potentially used for kHz repetition rate relativistic non-linear optics \cite{faure_review_2019,farace_multi-kilohertz_2026}. The simulations of section~\ref{sec:simBS} demonstrated that up-scaling of spectral broadening to the TW input peak power level is possible.  The major challenge is to find a good compromise between aperture and substrate thickness of the first beam splitter to minimize the B-integral in the substrate and to simultaneously avoid wavefront distortions by surface deformations of the beam splitter. However, all optical elements of the splitting and recombination units can be obtained with large aperture size, such that peak-power scaling requires mainly careful engineering of the setup.
	
	The most significant drawback of the divided pulse spectral broadening approach is the side-pulse generation which was observed for both the birefringent crystal and the intensity beam splitter approach. At FLASH, a contrast better than $10^3$ is sufficient in pump-probe experiments. However, a side pulse with 1-2\% relative intensity to the main pulse severely distorted the time-resolved experiments \cite{seidel_ultrafast_2022}. The achieved contrast with the birefringent crystal method is between $10^2$ and $10^3$, i.e. at the edge of the typical requirements for pump-probe experiments at the free-electron laser facility. The determined pre-pulse contrast on the order of $10^{-4}$ for the intensity splitter method is clearly not sufficient for relativistic optics experiments which typically require at least a $10^{-6}$ contrast to avoid pre-ionization of the target. Simulations showed that adding more beam splitters does not degrade the contrast further but rather adds additional side pulses. Nevertheless for relativistic nonlinear optics, an additional pulse cleaning stage becomes necessary. For instance, nonlinear ellipse rotation prior to dispersion compensation \cite{homoelle_pulse_2002,kalashnikov_characterization_2004,jullien_high-efficiency_2004}, which is essentially an intensity gating, can be very efficient and does not need much nonlinear phase shift for the initial better than $10^{-3}$ contrast. Moreover, the method is capable of handling high average powers. In both divided-pulse spectral broadening approaches, the contrast can potentially be further improved by optimizing the alignment for maximized recombination efficiency. For this purpose, machine-learning-based optimization algorithms \cite{tunnermann_deep_2019} could enable faster, more robust alignment and real-time efficiency maximization, particularly as the number of replicas increases.

	\section{Conclusion}
	We have demonstrated two passively stable architectures for temporal pulse division in MPC-based post-compression: a previously unexplored free-space scheme based on 50:50 intensity beam splitters, and a birefringent-crystal implementation used as an established benchmark. In both cases, recombination efficiency remains high while the broadened spectrum stays independent of replica number, confirming that pulse division decouples achievable pulse energy from the nonlinear dynamics governing spectral broadening, and that this decoupling holds regardless of which splitting mechanism is used. As discussed in Section~4, the two architectures are complementary rather than competing: the free-space scheme is particularly suited to peak power-scaling, free from the bulk-material limitations of birefringent crystals, while the crystal-based scheme offers established, compact repetition-rate flexibility without modifying the MPC itself. Together, these results establish passively stable pulse division as a practical route to extending the usable pulse energy of a fixed-geometry MPC, directly relevant to energy and repetition rate applications such as high-energy THz generation and laser-plasma acceleration.
	
	\section*{Acknowledgements}
	We thank DESY, a member of the Helmholtz Association HGF, for the provision of experimental facilities. Moreover, we thank the members of DESY's FS-LA and FS-PRI groups as well as the members of Clara Saraceno's group at Ruhr University Bochum for fruitful discussions.
	
	\section*{Funding}
	N.K. and M.S. acknowledge funding from the Federal Ministry of Research, Technology and Space (BMFTR, IGNYTe project, grant No. 13F1019)

	\section*{Author contributions}
	\textit{Experiments}: NJ, HS (lead); AS, MMA, NK, MS (supporting);
    \textit{Simulations}: MS (lead); NJ, GA (equal);
    \textit{Formal analysis}: NJ, HS, MS (lead); HT (supporting);
    \textit{Conceptualization}: MS (lead); NJ, HT, HC (equal);
    \textit{Resources}: IH (lead); VH (supporting);
    \textit{Supervision}: MS (lead); IH (supporting);
    \textit{Writing – original draft}: NJ (lead), MS (supporting);
    \textit{Writing – review and editing}: NJ, AS, NK, VH, GA, HC, MS (equal)

\section*{Data availability}
Data will be made available on reasonable request.

	\section*{Supplements}
	\addcontentsline{toc}{section}{Supplements}
	\renewcommand{\thesubsection}{S\arabic{subsection}}
	\setcounter{subsection}{0}

		\subsection{Numerical modeling of divided pulse spectral broadening with birefringent crystals}
		\label{suppl:Xtal_sim}
		\subsubsection{Methods.}
		\label{suppl:Xtal_methods}
		Modeling of the experiments with birefringent crystals was performed by simulating the nonlinear propagation through the MPC with the SISYFOS code \cite{arisholm_simulation_2021} and by multiplying the phase responses of the used optics to the complex electric fields which were propagated. Ideal Gaussian beams and pulses were assumed as input. A time grid with $2^{12}$ points and $3.5$\,fs time steps was used. Moreover, the spectral broadening simulation assumed cylindrical symmetry and used 256 radial grid points and a grid size that corresponded to five times the beam radius on the MPC mirrors. The spacing of the two Kerr plates was adjusted to 45\,mm in the simulations in order to match the numerical and experimental bandwidths after spectral broadening, and hence to adequately simulate the differential dispersion effects. To account for nonlinear ellipse rotation, nonlinear propagation was performed in the circular polarization basis and cross-phase modulation between left- and right-handed circularly polarized light was included. The waveplates were modeled by the retardance of true zeroth order quartz plates. In the same way, the delays introduced by the birefringent crystals were accounted for by adding the retardance $\Gamma = \pm \tfrac{1}{2} l_x (k_{sn}-k_{fn})$ to the spectral phases of the p/s polarized ultrashort pulses where $l_x$ is the birefringent crystal length and $k_{s(f)n}$ are the normal components of the wave vectors for the electric fields aligned along the slow (fast) crystal axis. Spatial walk-off caused by angle tuning of the fast axis' refractive index was neglected since the required tuning angle of the 10\,mm crystal was only about 1$^\circ$. For modeling the chirped and the HR mirrors, the theoretical GDD curves provided by the suppliers were interpolated to the frequency grid of the simulations. The compressor simulation numerically optimized the number of bounces from the chirped mirrors (Laseroptik 27089Gi1) for the highest output peak power.

	\subsubsection{Sources of losses and side-pulse generation}
		\label{suppl:Xtal_simresults}
		
		\begin{figure}[t!]
			\centering
			\includegraphics[width=1\linewidth]{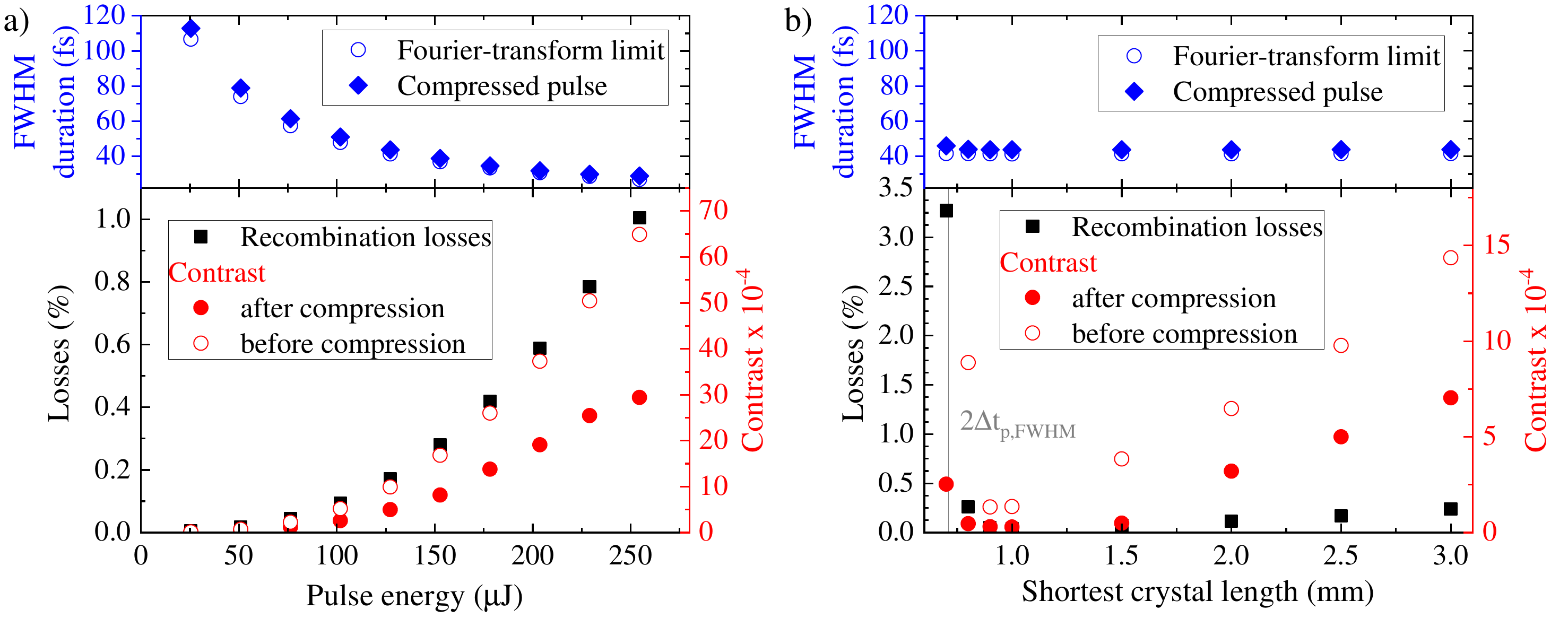}
			\caption{Variation of Fourier-transform limit, compressed pulse duration, recombination efficiency and side-to-main-pulse contrast in the dependence of (a) the total input pulse energy for divided pulse spectral broadening with the experimentally studied birefringent crystal sequence and of (b) the shortest splitting / combining crystal length at the experimental pulse intensities. As in the experiments, the second crystal is twice and the third four times as long as the shortest. }
			\label{fig:Xtal_en_len}
		\end{figure}
		
		\begin{figure}[b!]
			\centering
			\includegraphics[width=1\linewidth]{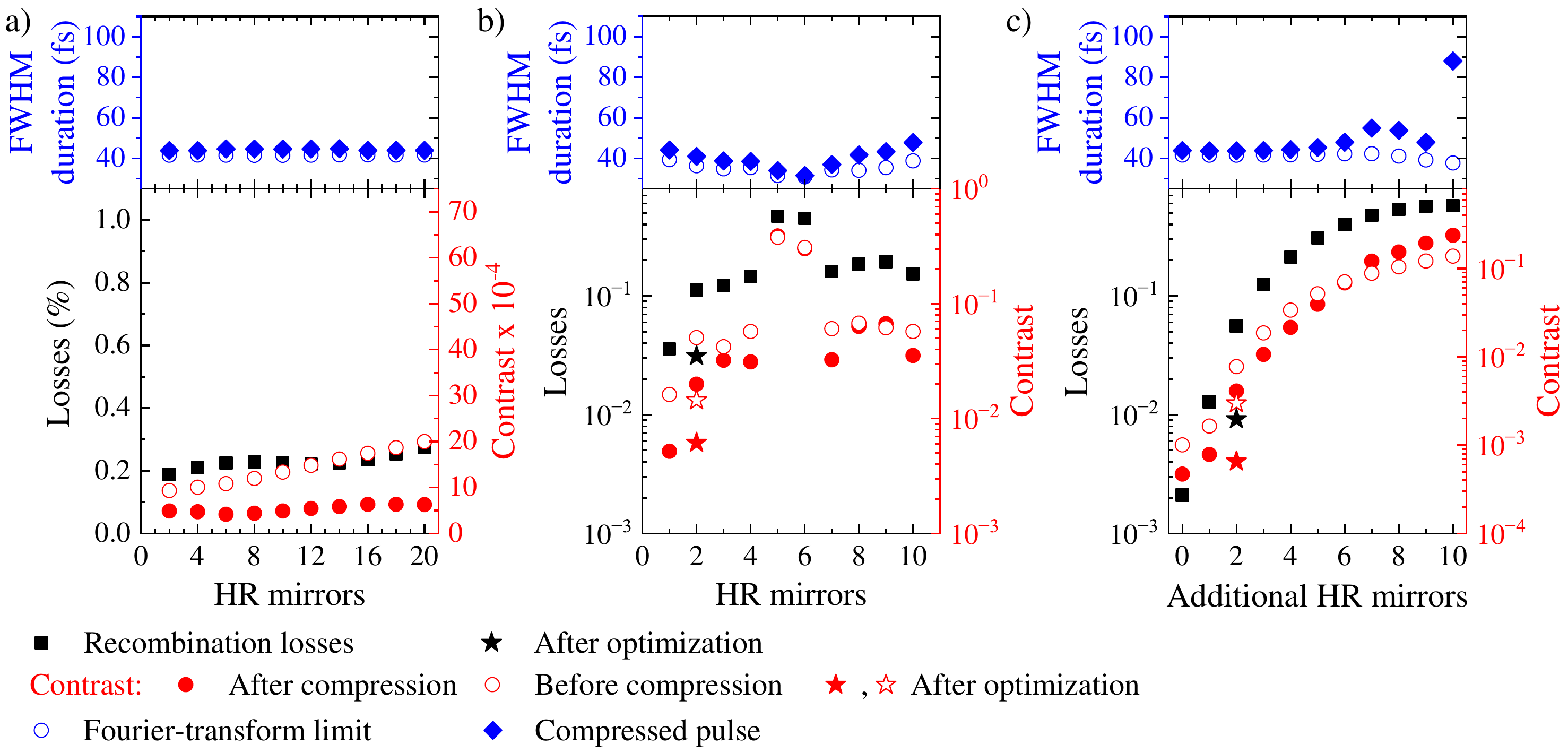}
			\caption{Variation of Fourier-transform limit, compressed pulse duration, recombination efficiency and side-to-main-pulse contrast in the dependence of (a) the number of HR mirrors between the splitting / recombining crystals and the QWP, of (b) the number of HR mirrors between the QWP and the MPC and of (c) the difference between number HR mirrors in the QWP-recombining-crystal-path and the splitting-crystal-QWP-path. The stars in (b) exhibit the improvement of efficiency and contrast after optimizing the QWP angle, the stars in (c) the improvement after tilting the longest crystal by about 1$^\circ$.}
			\label{fig:HR_mirror}
		\end{figure}
		
	Several simulation parameters were varied in order to investigate their impact on the recombination efficiency and the pulse contrast. First, the total pulse energy was varied between 25.5\,\textmu J and 255\,\textmu J. The resulting pulse durations, recombination losses and temporal contrasts are shown in Fig.~\ref{fig:Xtal_en_len}a. The pulses for the experimental operation point at 127\,\textmu J are plotted in Fig.~\ref{fig:Xtal_sim}a. The pulse energy emerges from dividing 140\,\textmu J by the 1.1 $M^2$-factor which is not considered in the simulations with fundamental Gaussian beams. In agreement with previous investigations \cite{guichard_analysis_2014,jacqmin_temporal_2015}, losses increase and contrast degrades with increasing pulse energy and compression factor, respectively. This is caused by the growing impact of differential GVD. Fig.~\ref{fig:Xtal_en_len}b studies the relation between input pulse duration and birefringent crystal length which was also addressed in reference \cite{jacqmin_temporal_2015}. As for the pulse energy scan, differential GVD causes recombination losses and contrast degradation which increase with crystal length. From that point of view, short crystals are advantageous. However, if the group delay difference between fast and slow crystal axes is less than 2.5-times the Gaussian FWHM pulse duration, a rapid increase of losses was observed being caused by spurious cross-phase modulation (XPM). For non-perfect Gaussian pulses, the minimal crystal length with negligible XPM losses might extend the least GD difference of 2.5-times the pulse duration.\par
		Further simulations investigated the impact of dielectric mirrors with 45$^\circ$ angle of incidence. The influence on the output pulses of mirrors at three different locations was studied: First, the common path mirrors between the splitting / combining crystals and the QWP; second, the common path mirrors between the QWP and the MPC; third, the additional mirrors in the path between the QWP and the combining crystals which were not in the path of the splitting crystals and the MPC (cf. Fig.~\ref{fig:xtal_setup}).  A matched number of dielectric mirrors between the birefringent crystals and the QWP has hardly an impact on recombination efficiency and contrast as Fig.~\ref{fig:HR_mirror}a demonstrates. The simulation study was performed with up to 20 mirrors, only four were in the experimental setup. Efficiency and contrast are most adversely affected by the oblique incidence HR mirrors between the QWP and the MPC. Even after optimizing the QWP angle, which was done numerically, sub-percent-level losses and a contrast better than $6\times 10^{-3}$ could not be recovered as shown in Fig.~\ref{fig:HR_mirror}b. A mismatch between HR mirrors in the paths between the birefingent crystals and QWP also leads to a rapid increase of losses and side-pulse amplitude. However, refractive index tuning of one birefringent crystal can theoretically reduce the losses to 0.9\,\% and yield a contrast of $6\times 10^{-4}$ for the experimental case of two additional HR mirrors in the path to the recombining crystals (Fig.~\ref{fig:HR_mirror}c).
		
\subsection{Numerical modeling of divided pulse spectral broadening with intensity beam splitters}
		\label{suppl:BS_sim}
		\subsubsection{Methods.}
		\label{suppl:BS_methods}
		The following modifications compared to the supplement \ref{suppl:Xtal_methods} were made: Due to the nanosecond delays, the nonlinear propagation of each pulse was simulated independently, i.e. eight SISYFOS simulations run for eight pulses (cf. Fig~\ref{fig:BS_sim}a) and 128 simulations for 128 pulses (cf. Fig~\ref{fig:BS_sim}b). For the eight pulse studies, the single simulation time grid contained $2^{9}$ points with $10$\,fs step size; for 128 pulse study it contained $2^{10}$ points with $5$\,fs step size. Due to the nanosecond delays, propagation effects in the splitting / combining units were considered. Like in the embedded SISYFOS simulations, the initial Gaussian was described by a 2D-array of complex numbers with the dimensions time and radius. To model spatial effects, the Hankel transform by means of the PyHank module was used \cite{guizar-sicairos_computation_2004}. Linear propagation through the splitting delay lines was done with a 8 to 16 times larger radial grid than the nonlinear propagation through the MPC. The step width was identical. The nonlinear simulation grid contained 256 radial points and extended over four times the mode-matched beam radius at the MPC mirrors. For simulating the recombination after spectral broadening of 128 pulses, the electric field was averaged over four radial grid points, i.e. a spatial grid reduction from 4096 to 1024 points was performed to avoid computer memory overflow. Test simulations with four beam splitters resulted in the same contrast and recombination efficiency with and without electric field averaging, confirming that averaging did not generate numerical artifacts. Mode-matching was generally simulated by a single lens with variable focal length. A $\pi$-phase shift was introduced to the s-polarized light after spectral broadening in order to allow recombination into a single pulse. In the experiment, a half-wave plate can be used for implementing this phase shift or a tiny adjustment in the the path difference between s- and p-polarized light.\par
		
\begin{table}[t]
			\caption{Impact of different optics in the setup and the collimated beam size in the splitting / recombining units on the side-to-main-pulse contrast.}
			\label{tab:contrast_optics}
			\centering
			\small
			\begin{tabular}{cccccccr}
				\toprule
				\textbf{QWP}$^1$ & \textbf{Medium}$^2$ & \textbf{Substrate}$^3$ & \textbf{\makecell{Beam \\ splitters$^4$}} & \textbf{TFP}$^5$ & \textbf{\makecell{45$^\circ$ HR\\ mirrors$^6$}} & \textbf{\makecell{Collimated \\ beam radius$^7$}} & \textbf{Contrast} \\
				\midrule
				ideal & vacuum & none & ideal & ideal & 0 & 3\,mm & $4.4 \times 10^{-6}$ \\
				\textbf{real} & vacuum & none & ideal & ideal & 0 & 3\,mm & $6.1 \times 10^{-6}$ \\
				ideal & \textbf{air} & none & ideal & ideal & 0 & 3\,mm & $4.8 \times 10^{-6}$ \\
				ideal & vacuum & \textbf{silica} & ideal & ideal & 0 & 3\,mm & $4.4 \times 10^{-6}$ \\
				ideal & vacuum & none & \textbf{real} & ideal & 0 & 3\,mm & $1.2 \times 10^{-5}$ \\
				ideal & vacuum & none & ideal & \textbf{real} & 0 & 3\,mm & $5.7 \times 10^{-5}$ \\
				\textbf{real} & vacuum & none & ideal & ideal & \textbf{1} & 3\,mm & $1.6 \times 10^{-4}$ \\
				\textbf{real} & \textbf{air} & \textbf{silica} & \textbf{real} & \textbf{real} & \textbf{1} & 3\,mm & $2.1 \times 10^{-4}$ $^*$ \\
				\midrule
				ideal & vacuum & none & ideal & ideal & 0 & \textbf{8\,mm} & $1.0 \times 10^{-7}$ \\
				ideal & vacuum & none & ideal & ideal & 0 & \textbf{4\,mm} & $1.5 \times 10^{-6}$ \\
				ideal & vacuum & none & ideal & ideal & 0 & \textbf{2\,mm} & $2.1 \times 10^{-5}$ \\
				ideal & vacuum & none & ideal & ideal & 0 & \textbf{1\,mm} & $1.9 \times 10^{-4}$ \\
				\bottomrule
			\end{tabular}
			\caption*{$^1$ Ideal means perfect circular polarization in the MPC and perfect linear polarization in the recombination unit. Real means that the polarization change emerges from a true zeroth order quartz waveplate. $^2$ The medium in the delay lines leads to differential GVD in short and long paths. $^3$ Only the beam splitter substrate is considered with 4\,mm thickness, the TFP substrate is not considered. $^4$ Ideal beam splitters exhibit 50\,\% reflectance and zero GDD, real ones take the theoretical data provided by the supplier. $^5$ For the real TFP, the theoretical coating data of 56$^\circ$ angle of incidence beam splitters is taken (EKSMA 420-1268E). $^6$ Applies same GD difference like in study of supplement~\ref{suppl:BS_simresults}. The QWP angle is numerically rotated to optimize the contrast. $^7$ A 3\,mm radius yields a 27\,m Rayleigh length, 4.5\,ns delay corresponds to an optical path 1.35\,m in vacuum. $^*$ Settings of this row lead to pulse train shown in Fig.~\ref{fig:BS_sim}a}
		\end{table}		

\subsubsection{Sources of side-pulse generation}
		\label{suppl:BS_simresults} 
		
		The numerical side-pulse studies mimicked the experimental setup described in section \ref{sec:ExpBS} with small modifications: FROG measurements of the compressed pulses from the Yb:YAG amplifier revealed a multi-ps long pedestal under the main pulse which effectively reduced the peak power of the source compared to a pedestal-free pulse with the same FWHM. The simulations with Gaussian pulses accounted for this by a 28\,\% pulse energy reduction. Consequently, the experimentally and numerically obtained Fourier-transform limits were comparable. The simulation of the chirped mirror compressor after recombination yielded pulses with 85\,fs duration. To investigate the impact of the used optics on the recombination losses and side pulse generation, the refractive index of air (Sellmeier equation from refs.~\cite{ciddor_refractive_1996,polyanskiy_refractiveindexinfo_2024}) in the long delay lines, the dispersion of the beam splitter substrates \cite{polyanskiy_refractiveindexinfo_2024,malitson_interspecimen_1965}, the theoretical reflectance and transmission curves of the beam splitters and the TFP as well as the theoretical GDD curves of the beam splitters were considered. Data for the anti-reflection coated back surface of the beam splitters and the GDD of the TFP was not available. Furthermore, as for the birefringent crystal splitting approach, the impact of 45$^\circ$ HR mirrors between the QWP and the MPC was considered. In this case, only one mirror was in the setup. The upper part of Table~\ref{tab:contrast_optics} shows how the optical elements contribute to the side-to-main-pulse contrast of the simulated pulse trains. Like in the study presented in supplement~\ref{suppl:Xtal_simresults}, the 45$^\circ$ HR mirror, which was the pick-off mirror in our setup (Fig.~\ref{Fig 2.}), was the main source of contrast degradation. It limits the contrast to the $10^{-4}$-order which was also reached in our experiments. By using a hole in one of the MPC mirror substrates for in- and out-coupling, HR mirrors between QWP and MPC can be omitted. The polarization contrast of the TFP was the second strongest source for side-pulse generation. The simulation did not include the s-polarized light transmitted through and the p-polarized light reflected from the polarizers. First, reflectance and transmittance of the beam splitters for p-polarized light was not known, and thus modeling could have only been based on assumptions. Second, the FROG measurements are also polarization sensitive and cannot detect side pulses with opposite polarization. Beam splitters can be manufactured with nearly perfect 50:50 transmittance/reflectance ratio and very small GDD. Therefore, the partially reflecting surfaces of the beam splitters led only to side-pulses with $1\times 10^{-5}$ relative intensity. The anti-reflection coated back surfaces, which could not be modelled, probably generate stronger side pulses. Slightly wedged, precisely oriented substrates may prevent this side-pulse generation at the expense of small spatial chirp. In contrast to the birefrigent crystal studies, differential GDD hardly degrades the side-to-main-pulse contrast. Both the inclusion of air and of the silica substrate dispersion kept the contrast on the $10^{-6}$-order.  \par
		The lower part of Table~\ref{tab:contrast_optics} shows how the collimated beam size in the splitting / combination unit influences the side-pulse amplitude. The meter-scale path differences lead to sligthly different beam sizes in the MPC and could inhibit complete recombination after spectral broadening. Our study shows, however, that the beam size and wavefront variation of the splitted pulses makes only a minor contribution to the side-to-main-pulse contrast. The experimental beam radii were 2.6\,mm in x- and 3\,mm in y-directions. According to Table~\ref{tab:contrast_optics}, the propagation effect would only lead to relative side-pulse intensities on the high-$10^{-6}$- to low-$10^{-5}$-level, i.e. the used optics are the dominating sources for contrast degradation. Only for 1\,mm input radii and Rayleigh lengths which approach the path lengths in the delay lines, the contrast reaches the $10^{-4}$-level. Input peak power scaling requires the use of larger beam sizes and consequently larger optics, and thus longer delay lines can also be used without significant side-pulse generation. Fig.~\ref{fig:BS_sim}b shows that a side-to-main-pulse contrast on the order of $10^{-5}$ despite a 52\,ns delay between longest and shortest path in the splitting unit.
		
		\subsubsection{Delay lines for high-energy pulse compression simulation.}
		\label{suppl:delay_lines}
		
		\begin{figure}[!b]
			\centering
			\includegraphics[width=1\linewidth]{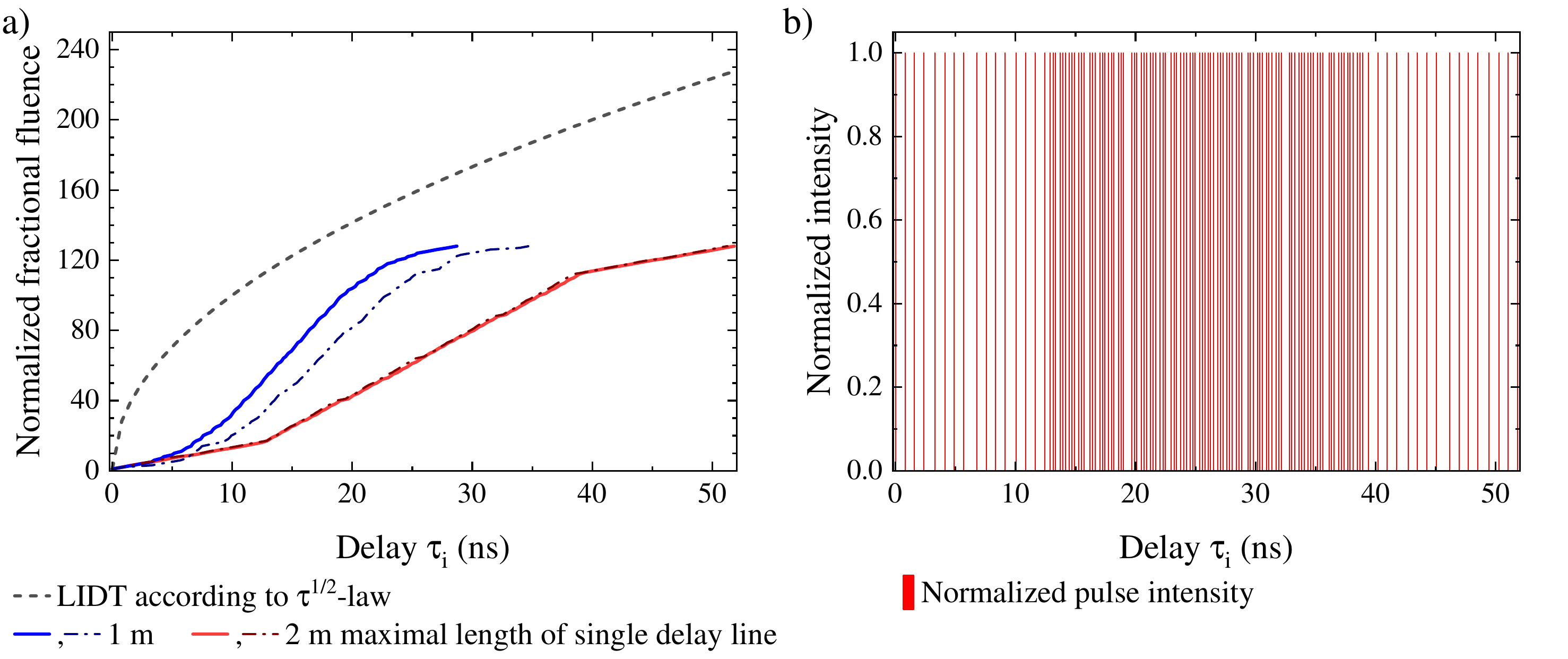}
			\caption{(a) Delay-dependent fractional fluence for different delay line lengths compared to the expected laser damage threshold (LIDT) fluence. Solid lines: maximization of the $\Delta \tau / (\tau\Delta t_p)^{{1}/{2}}$ minimum, dashed-dotted lines: maximization of the $\Delta \tau_\text{min}$. (b) Pulse train generated by the splitting unit for simulating the nonlinear compression of a 1\,TW peak power pulse. The minimal delay between two pulses is $\Delta \tau_\text{min} = 205\,ps$. }
			\label{fig:pulse_train}
		\end{figure}
		
		This 52\,ns delay emerged from setting lower and upper limits for the delay paths. The lower limit is given by the beam splitter size and the beam's angle of incidence $\alpha$. For $d_M = 6"$ optics, the minimal displacement of the delay line HR mirrors is $d_\text{min}= d_M \sin(\alpha) \approx 108$\,mm. Otherwise, the beam would get clipped. For equidistant delays, the maximal displacement would be $d_\text{max}= d_\text{min}\cdot 2^{N_d-1}$, i.e. 6.9\,m for $N_d = 7$ beam splitters, which is impractical. However, equidistant delays are neither necessary nor ideal, because the laser induced damage threshold (LIDT) fluence is expected to scale by $F_\text{LIDT}\propto(\tau/\Delta t_p)^{1/2}$ \cite{ristau_laser-induced_2014} where $\tau$ is the pulse train duration and $\Delta t_p$ the single pulse duration. For damage threshold investigations, we define the fractional fluence as the sum of single pulse fluences of the pulses in the interval [0, $\tau_i$] where $\tau_i \leq \tau$ is the delay of the i$^\text{th}$ divided pulse. For equidistant delays, the fractional fluence would scale linearly with $\tau_i$. To find a suited set of delay lines, we first fixed the upper limit. For the simulation results presented in Fig.~\ref{fig:BS_sim}b, a maximal displacement (that is L in Fig.~\ref{Concept}c) of 2\,m was allowed which corresponded by choice to the MPC length. The lower displacement limit was set to 12\,cm which is slightly longer than permitted by the beam splitters. The other five delays were computed by running a random number generator for 10\,minutes and keeping the optimal set of delays. One optimization was maximizing the smallest delay between to consecutive pulses $\Delta \tau_\text{min}$. The other optimization was maximizing the smallest slope ratio between theoretical damage threshold and computed change of fluence $(dF_\text{LIDT}/dt) / (\Delta F/ \Delta \tau)$ which effectively means optimizing $\Delta \tau / (\tau_i\Delta t_p)^{1/2}$. For 2\,m maximal length, both optimizations yielded almost the same pulse distribution as shown in Fig.~\ref{fig:pulse_train}a. The corresponding sequence of the 128 pulses is shown in Fig.~\ref{fig:pulse_train}b. The fractional fluence stays for all delays clearly below the expected LIDT, and thus the damage threshold is determined by the single pulse like for MPC spectral broadening without pulse division. This is even valid for shorter maximal delays and seven beam splitters. Fig.~\ref{fig:pulse_train}a also shows the fractional fluence evolution for a maximal displacement of 1 m. In this case, the optimization strategies yielded different results: For maximizing $\Delta \tau_\text{min}$, the difference between $F_\text{LIDT}$ and the pulse train fluence is larger for every delay. The optimization of $\Delta \tau / (\tau\Delta t_p)^{1/2}$ yielded a shorter total delay. For the 6" beam splitters, the collimated input beam radius was set to $w_\text{in}= d_M \sin(\alpha)/4 = 26.94$\,mm. Despite the large sizes, a 1\,TW pulse would still acquire a $\pi/10$ B-integral in the first beam splitter after 6.6\,mm propagation through its substrate. We have not investigated the impact of B-integral in the beam splitters numerically. As spectral broadening in gas-filled MPCs does not change much the pulse shape, even a moderate B-integral accumulation might be tolerable. However, a quantitative study on the expected losses and the consequences for the side-to-main-pulse contrast has not been performed. In any case, proper engineering, in particular of the first beam splitter, will be major challenge for the high-energy pulse compression implementation.

		\subsection{Third Order Autocorrelation device.}
		\label{suppl:TOAC}
\begin{figure}[!h]
			\centering
			\includegraphics[width=0.7\linewidth]{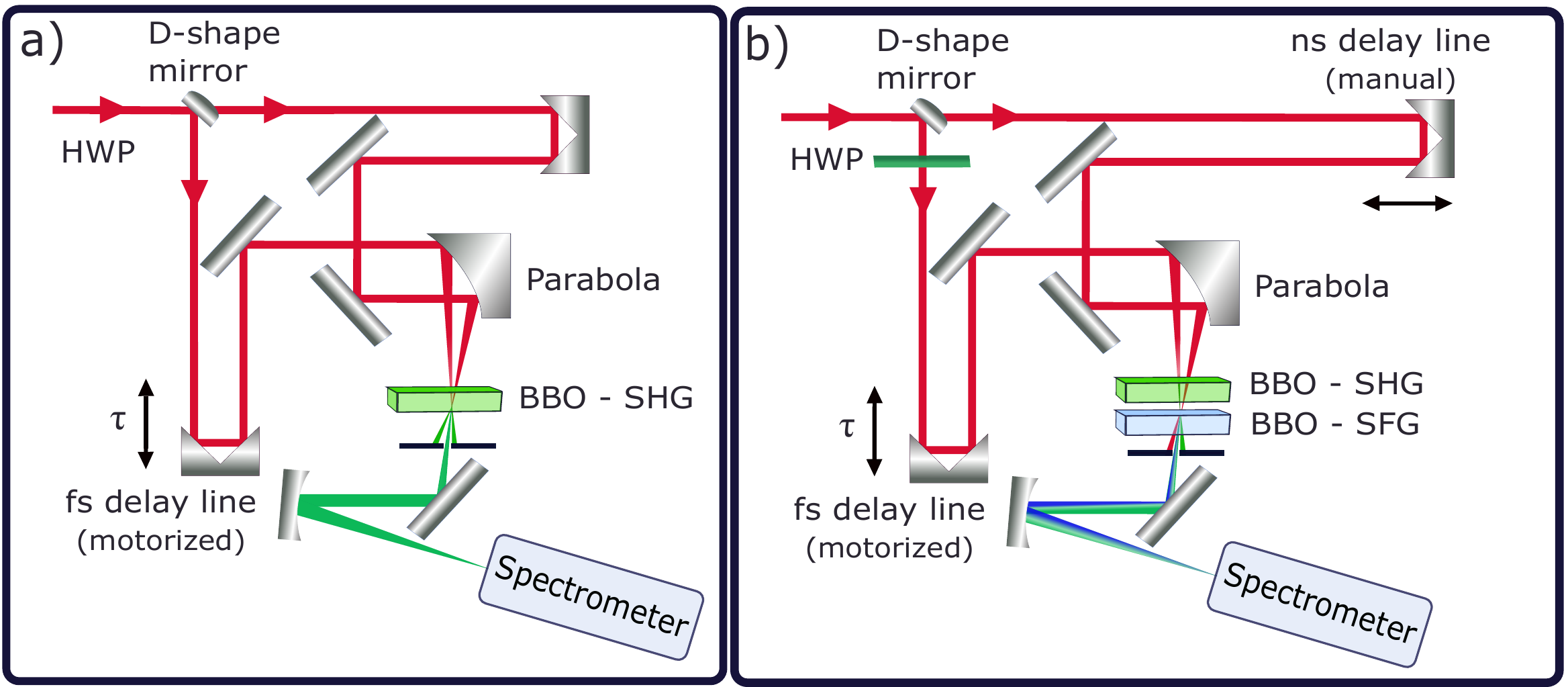}
			\caption{\textbf{Schematic of (a) the SHG-FROG and (b) the third-order autocorrelator.} In (b), a long-travel ns delay stage is added to arm~2 and the signal is generated in a cascaded BBO geometry (Type~I SHG and SFG) to produce $3\omega$. The HWP is shown in the shorter delay line. This means that the SH precedes the fundamental, and thus pre-pulses are detected.}
			\label{fig:xfrog_setup}
		\end{figure}

		For the pulse duration measurements presented in Figs.~\ref{Crystals_frog}, \ref{Side_pulses_crystals} and \ref{BS_frog}, the FROG setup shown in Fig.~\ref{fig:xfrog_setup}a and explained in ref.~\cite{seidel_multi-watt_2018} was used. To evaluate the temporal contrast over a nanosecond delay range and to eliminate the directional ambiguity inherent to standard second-harmonic generation (SHG), this home-built FROG device was modified to work like a third-order autocorrelator (Fig.~\ref{fig:xfrog_setup}b). Commercial third-order autocorrelators are difficult to operate with the burst-mode laser we used \cite{viotti_temporal_2021}.
		
		The incoming laser beam was spatially split into two independent arms using a D-shaped mirror. One arm contained a fine motorized optical delay line ($\tau$) for high-resolution temporal scanning, while the second arm incorporated a long-travel manual stage (ns delay line) to extend the measurement window into the nanosecond regime. The manual stage was explicitly set to the specific temporal positions where the side pulses were expected to appear, followed by a precision scan with the motorized delay line.
		
		A third-order non-collinear interaction $\chi^{(3)}$ was achieved by focusing both beams with an off-axis parabolic mirror into a cascaded two-crystal geometry of $\beta$-barium borate (BBO) crystals, $\text{BBO}_\text{SHG}$ and $\text{BBO}_\text{SFG}$. In the first crystal, $\text{BBO}_\text{SHG}$, one of the arms undergoes second-harmonic generation (SHG) to generate a $2\omega$ reference gate. The other beam has orthogonal polarization, and thus was not frequency-converted. In the second crystal, $\text{BBO}_\text{SFG}$, this $2\omega$ gate was nonlinearly mixed with the fundamental $\omega$ pulse from the complementary arm via type I sum-frequency generation (SFG, $\omega + 2\omega \rightarrow 3\omega$). The resulting $3\omega$ signal was spatially isolated by an iris, imaged, and recorded using a high-dynamic-range spectrometer.
		
		To selectively measure pre-pulses, the HWP was placed in the arm with the motorized fs delay line, to measure post-pulses the HWP was placed in the arm with the ns delay line. Polarization rotation by the HWP controlled which arm generated the $2\omega$ gate pulse in $\text{BBO}_\text{SHG}$ and thereby defined the temporal reference direction relative to the pulse under test.
		
\printbibliography
\end{document}